\documentclass{pasj00}
\twocolumn

\newcommand{\tautot}{{\tau_{d,\rm eff}^{\rm tot}}}
\newcommand{\hi}{{\rm H\emissiontype{I}} }
\newcommand{\hins}{{\rm H\emissiontype{I}}}

\begin{document}
\SetRunningHead{Totani et al.}{Infrared SEDs of Galaxies in the AKARI
All Sky Survey}
\Received{2011/03/28}
\Accepted{2011/06/18}
\Published{2011/12/25}

\title{Infrared Spectral Energy Distribution of Galaxies
in the AKARI All Sky Survey: Correlations with Galaxy Properties,               
and Their Physical Origin}



%
 \author{%
   Tomonori \textsc{Totani}\altaffilmark{1},
   Tsutomu T. \textsc{Takeuchi}\altaffilmark{2,3},
   Masahiro \textsc{Nagashima}\altaffilmark{4},
   Masakazu A.R. \textsc{Kobayashi}\altaffilmark{5}, 
   \\ and
   Ryu \textsc{Makiya}\altaffilmark{1}}
 \altaffiltext{1}{Department of Astronomy, School of Science,
    Kyoto University, Sakyo-ku, Kyoto 606-8502}
  \altaffiltext{2}{Institute for Advanced Research, Nagoya University,
    Furo-cho, Chikusa-ku, Nagoya 464--8601}
  \altaffiltext{3}{Department of Particle and Astrophysical Science,
    Nagoya University, Furo-cho, Chikusa-ku, Nagoya 464--8602}
  \altaffiltext{4}{Faculty of Education, Nagasaki University,
    Nagasaki, Nagasaki 852-8521} 
  \altaffiltext{5}{Optical and Infrared
    Astronomy Division, National Astronomical Observatory, Mitaka,
    Tokyo 181-8588}


\maketitle

\begin{abstract}
  We have studied the properties of more than 1600 low-redshift
  galaxies by utilizing high-quality infrared flux measurements of the
  AKARI All-Sky Survey and physical quantities based on optical and
  21-cm observations. Our goal is to understand the physics
  determining the infrared spectral energy distribution (SED). The
  ratio of the total infrared luminosity $L_{\rm TIR}$, to the
  star-formation rate (SFR) is tightly correlated by a power-law to
  specific SFR (SSFR), and $L_{\rm TIR}$ is a good SFR indicator only
  for galaxies with the largest SSFR. We discovered a tight linear
  correlation for normal galaxies between the radiation field strength
  of dust heating, estimated by infrared SED fits ($U_h$), and that of
  galactic-scale infrared emission ($U_{\rm TIR} \propto L_{\rm
    TIR}/R^2$), where $R$ is the optical size of a galaxy. The
  dispersion of $U_h$ along this relation is 0.3 dex, corresponding to
  13\% dispersion in the dust temperature. This scaling and the
  $U_h/U_{\rm TIR}$ ratio can be explained physically by a thin layer
  of heating sources embedded in a thicker, optically-thick dust
  screen. The data also indicate that the heated fraction of the total
  dust mass is anti-correlated to the dust column density, supporting
  this interpretation. In the large $U_{\rm TIR}$ limit, the data of
  circumnuclear starbursts indicate the existence of an upper limit on
  $U_h$, corresponding to the maximum SFR per gas mass of $\sim 10 \rm
  \ Gyr^{-1}$.  We find that the number of galaxies sharply drops when
  they become optically thin against dust-heating radiation,
  suggesting that a feedback process to galaxy formation (likely by
  the photoelectric heating) is working when dust-heating radiation is
  not self-shielded on a galactic scale. Implications are discussed
  for the $M_{\hi}$-size relation, the Kennicutt-Schmidt relation,
  and galaxy formation in the cosmological context.
\end{abstract}

\section{Introduction}
\label{section:intro}

Dust grains in interstellar medium (ISM) absorb ultraviolet and
optical photons emitted from stars, and the energy is re-emitted from
dust grains in infrared and submillimeter wavelengths.  The spectral
energy distribution (SED) of this dust emission has a broad peak of
the modified blackbody in far-infrared bands from large grains in
thermal equilibrium with temperatures of $T_d \sim$ 10--50 K. The SED
in near- to mid-infrared bands is, on the other hand, characterized by
emissions from small grains and polycyclic aromatic hydrocarbons
(PAHs) that have a wide distribution of temperature because of
stochastic single-photon heating (Desert et al. 1990; see Draine 2003
for a review). Physical dust models about compositions and size
distribution have been developed to calculate SEDs of dust emission
for a given heating radiation field strength, $U$, and one can predict
the global SED integrated over a galaxy by summing up the emission
from dust with various values of $U$. Such models have been applied to
fit observed SEDs of nearby galaxies to derive physical quantities
(e.g., Dale et al. 2001; Dale \& Helou 2002; Draine et al. 2007; da
Cunha et al. 2008, 2010; Mu\~noz-Mateos et al. 2009; Ma{\l}ek et
al. 2010).

The intensity of the cosmic infrared background radiation indicates
that about half of the energy originally radiated from stars is
eventually re-emitted from dust grains (Hauser \& Dwek 2001). Heavily
obscured star formation activity is difficult to detect by
observations in ultraviolet/optical observations, and
infrared/submillimeter observations should crucially be important to
get the full picture of the formation and evolution of galaxies.  In
the last decade, our observational knowledge about the hidden side of
galaxy evolution has significantly improved thanks to the advanced
infrared/submillimeter facilities such as JCMT/SCUBA, Spitzer, AKARI,
and Herschel.  Moreover, the Atacama Large Millimeter/submillimeter
Array (ALMA) will revolutionize our understanding of galaxy evolution
in the near future.  In order to extract useful information from these
data, we need good theoretical understanding and models of galactic
dust emission to be compared.

However, it is not straightforward to predict global (i.e.,
galactic-scale) SEDs of dust emission for a theoretical model galaxy,
in contrast to direct stellar emission that can be calculated by the
method of stellar population synthesis.  This is because the physics
determining the global SED of dust emission and its relations to other
physical properties have remained highly elusive (see Walcher et
al. 2011 for a review).  An approach often taken in the literature
(e.g., Guiderdoni et al. 1998; Takeuchi et al. 2001; Chary \& Elbaz
2001; Lagache, Dole, \& Puget 2003; Valiante et al. 2009) is to relate
the SED parameter (e.g., the dust temperature $T_d$) to the total
infrared luminosity, $L_{\rm TIR}$ (defined as the bolometric
luminosity of dust emission), based on the observed correlation
between the two (higher $T_d$ for larger $L_{\rm TIR}$; Soifer et
al. 1987a; Soifer \& Neugebauer 1991; Chapin et al. 2009).  However,
there is a large scatter along the mean $L_{\rm TIR}$-$T_d$ relation
(see, e.g., Hwang et al. 2010 for recent data), and it is physically
unreasonable to relate an extensive quantity $L_{\rm TIR}$ that scales
with the system size to an intensive quantity $T_d$ that does not
(Totani \& Takeuchi 2002). Extension of this relation including
another parameter has also been discussed, e.g., galaxy size (Devereux
1987; Phillips \& Disney 1988; Lehnert \& Heckman 1996; Chanial et
al. 2007; Rujopakarn et al. 2011a,b), but a consistent physical
picture has not yet been established.

To make a completely theoretical prediction, one must solve transfer
of dust heating radiation taking into account complicated geometry in
a galaxy. Infrared SED models by such {\it ab initio} approaches have
also been developed (Silva et al. 1998; Takagi et al. 2003a, b; Dopita
et al. 2005), and applied to predict infrared SEDs in galaxy evolution
models (Granato et al. 2000; Baugh et al. 2005; Fontanot et
al. 2007). However, it is difficult to calculate detailed geometry
within galaxies in models on the cosmological scale, and one must add
many uncertain and adjustable model parameters, making it difficult to
extract useful information from comparison with observed data.

Therefore it is still important to investigate the relation between
infrared SEDs and various physical properties to find any correlation
that would be useful to better understand the galactic-scale radiation
from dust grains. The recently released catalogs of the AKARI All Sky
Survey (Ishihara et al. 2010; Yamamura et al. 2010) for sources
detected by the AKARI satellite (Murakami et al. 2007) provide us with
a new opportunity to investigate nearby galaxies for this purpose. The
Infrared Camera (IRC, Onaka et al.  2007) and the Far-Infrared
Surveyer (FIS, Kawada et al. 2007) on board the satellite have
detected more than $800 \, 000$ and $400 \, 000$ sources in 9--18 and
65--160 $\mu$m bands, respectively.  For recent studies on galaxies
using samples based on these catalogs, see Takeuchi et al. (2010),
Buat et al. (2011), Goto et al. (2011a, b), and Yuan et al. (2011).

Especially, the four FIS bands at 65--160 $\mu$m should be very useful
to accurately measure the modified blackbody peak of the dust emission
SEDs, as they cover the wavelength range around the peak by a larger
number of photometric bands with better flux sensitivity and angular
resolution than those of the previous infrared all-sky survey by the
Infrared Astronomical Satellite (IRAS, Soifer et al.  1987b).  In this
paper we perform SED fittings of physical dust models to the AKARI
data of more than 1600 low-redshift galaxies, and compare the SED
parameters with other physical properties of galaxies obtained by
optical and 21-cm observations.  The goal of this paper is to find the
key physical quantities and laws to determine the global SEDs of dust
emission (especially around the thermal peak), and give physical
interpretations to them.

In \S\ref{section:sample} we describe the construction of the galaxy
samples used in this work.  The physical dust models and SED fittings
to the AKARI data are described in \S\ref{section:SED-fit}.  We then
present theoretical background and considerations in
\S\ref{section:theory}, which will be used to interpret the results
obtained in this work.  The main results are presented in
\S\ref{section:result}, followed by discussions in
\S\ref{section:discussion} and summary in \S\ref{section:summary}.  We
adopt the standard cosmological parameters of $H_0 = 70 \ \rm km \:
s^{-1} \: Mpc^{-1}$, $\Omega_M = 0.3$, and $\Omega_\Lambda = 0.7$,
though the choice of the cosmological parameters hardly affects our
results based on galaxies at low redshifts.

\section{The Samples}
\label{section:sample}

\subsection{The AKARI All-Sky Survey}

\begin{figure}
  \begin{center}
    \includegraphics[width=6.3cm,angle=-90,scale=0.95]{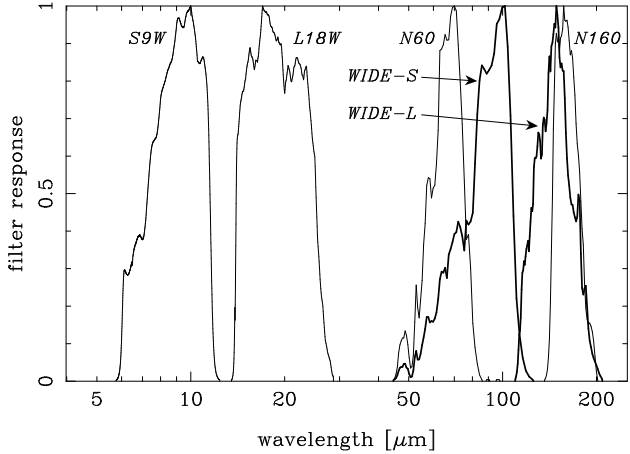}
  \end{center}
  \caption{The relative spectral response function (RSRF) against
    energy flux for the two AKARI/IRC bands ({\it S9W} and {\it L18W})
    and the four AKARI/FIS bands ({\it N60}, {\it WIDE-S}, {\it
      WIDE-L}, and {\it N160}).  The RSRFs are normalized by their
    maximum values.  }
  \label{fig:filters}
\end{figure}

We use the AKARI/IRC All-Sky Survey Point Source Catalog (PSC,
ver. 1.0) and the AKARI/FIS All-Sky Survey Bright Source Catalog (BSC,
ver. 1.0), which were released in March 2010 including $870 \: 973$
and $427 \: 071$ sources, respectively.  The IRC provides fluxes in
two photometric bands, {\it S9W} and {\it L18W}, whose central
wavelengths ($\lambda_c$) are 9 and 18 $\mu$m, respectively.  The FIS
provides fluxes in the four photometric bands of {\it N60}, {\it
  WIDE-S}, {\it WIDE-L}, and {\it N160}, whose central wavelengths are
65, 90, 140, and 160 $\mu$m, respectively.  The sensitivity limits
corresponding to 50\% detection completeness are shown in Table
\ref{tab:akari}, and the relative spectral response functions (RSRFs)
for the six bands are shown in Fig. \ref{fig:filters}.  See the
release notes of these catalogs available on the AKARI web site for
more details.

Since the SED fitting is the central issue of this work, treatment of
flux errors is important. In this work we consider only statistical
flux errors associated with each flux measurements, and systematic
errors that are common for all sources (e.g., flux calibration errors)
are not taken into account. Systematic errors would result in
systematic bias in SED fit parameters (e.g., a systematic shift of
dust temperature), but it does not seriously affect the study of
correlation or trend between the fit parameters and other galaxy
properties. Statistical errors can be divided into two parts: one is
the error that does not depend on source fluxes (denoted as $\Delta
F_{\rm const}$), and the other is the error that is proportional to
flux ($\Delta F = \epsilon F$).  The total flux error is assigned for
a flux measurement in a band as:
\begin{eqnarray}
  \Delta F_{\rm tot} = \left[ \: (\Delta F_{\rm const})^2 
  + (\epsilon F)^2 \: \right]^{\frac{1}{2}} \ .
\end{eqnarray}
For faint sources $\Delta F_{\rm const}$ is dominant while $\epsilon
F$ is dominant for bright sources. The 1$\sigma$ values of $\Delta
F_{\rm const}$ and $\epsilon$ for each band are estimated from the
signal-to-noise distributions presented in the release notes of the
AKARI catalogs, and they are summarized in Table \ref{tab:akari}.

\begin{table*}
  \caption{Summary of the AKARI All-Sky Survey Parameters}
  \label{tab:akari}
  \footnotesize
  \begin{center}
    \begin{tabular}{lccccccc}
      \hline
      instrument & \multicolumn{2}{c}{IRC} & & \multicolumn{4}{c}{FIS} \\ 
      \cline{2-3} \cline{5-8}
      band name & {\it S9W} & {\it L18W} & & {\it N60} & {\it WIDE-S} 
         & {\it WIDE-L} & {\it N160} \\
      \hline
      central wavelength $\lambda_c$ [$\mu$m] 
         & 9 & 18 & & 65 & 90 & 140 & 160 \\
      flux sensitivity$^*$ [Jy] & 0.09 & 0.17 & & 2.6 & 0.39 & 2.9 & 5.9 \\
      $\Delta F_{\rm const}$ [Jy] & 0.02 & 0.04 & & 0.48 & 0.11 & 0.28 & 1.26 \\
      $\epsilon$ [\%] & 5.6 & 6.7 & & 10 & 10 & 10 & 10 \\
      PSF [arcsec FWHM] & 5.6 & 5.6 & & 37 & 39 & 58 & 61 \\ 
      \hline 
    \end{tabular}
    \begin{minipage}{0.6\hsize}
      { \footnotesize
        $^*$The flux where the source 
        detection completeness is 50\%. }
    \end{minipage}
  \end{center}
\end{table*}

\subsection{Cross-Matches with Optical and 21cm Data}

We construct the following three samples by cross-matching the AKARI
catalogs with the existing catalogs of nearby galaxies selected in
other wavelengths.  The first sample is the cross-match with the
galaxies detected by the Sloan Degital Sky Survey (SDSS), and the
second with the \hi Parkes All-Sky Survey (HIPASS) selected by
21-cm flux.  We regard these as the two main samples in this paper,
and we can check the dependence on the sample selection by these two.
In addition, we construct the third sample to extend the dynamic range
of surface density of star formation rate (SFR), which is the
cross-match with the Kennicutt (1998, hereafter K98) sample of
nearby circumnuclear starburst (SB) galaxies.
The summary of constructing these samples, showing 
galaxy numbers at various stages of selections, is presented
in Table \ref{table:sample-summary}.

\begin{table*}
  \caption{The Summary of Sample Construction}
  \label{table:sample-summary}
  \footnotesize
  \begin{center}
    \begin{tabular}{lrrr}
      \hline
       & AKARI-SDSS & AKARI-HIPASS & AKARI-SB \\
      \hline
      original catalog     & 1145617       &    4315     &       36 \\
      match with AKARI/FIS &    6500       &    1527     &       34 \\
      FIS flux quality flag &   1138       &     840     &       33 \\
      distance cut$^*$      &    996       &     711     &       32 \\
      magnitude \& size cut$^\dagger$ &    878       &  -- \hspace{0cm} & 31 \\
      $L_{\rm TIR, K98}/L_{\rm TIR} \ge 0.5 $ & -- \hspace{0cm} &
           -- \hspace{0cm}  &  24 \\
      \hline
      final sample &             878       &     711     &       24 \\
      \hline
      stellar mass$\ddagger$       &    550       &       0     &        0 \\
      star formation rate$\ddagger$ &    567       &       0     &        0 \\
      metallicity$\ddagger$        &    246       &       0     &        0 \\
      $\tau_V$              &    534       &       0     &        0 \\
      \hi gas mass          &    257       &     711     &        0 \\
      H$_2$ gas mass        &      0       &       0     &       24 \\
      \hline
    \end{tabular}
    \begin{minipage}{0.8\hsize}
      Notes. --- The upper six rows show the numbers of galaxies after
      applying certain selection criteria in the process of
      constructing the samples. The lower six rows show the numbers of
      galaxies for which a certain
      physical quantity is available. \\
      $^*$The condition of $z \ge 0.005$ for AKARI-SDSS and
      AKARI-HIPASS, while $z \ge 0.005$ or availability of
      redshift-independent distance is required for AKARI-SB. \\
      $^\dagger$The condition of $r \le 16.0$ mag and $\theta_{P90}
      \ge 6''$ is adopted for AKARI-SDSS, while a galaxy without
      diameter size
      in the K98 sample is excluded from AKARI-SB. \\
      $\ddagger$Physical quantities derived from SDSS optical spectra.
    \end{minipage}
  \end{center}
\end{table*}

\vspace{0.2cm}
\subsubsection{The AKARI-SDSS Sample}

We first extracted a sample of $1 \: 145 \: 617$ SDSS galaxies having
measured redshifts from the DR7 NYU Value-Added Galaxy Catalog
(NYU-VAGC, Blanton et al. 2005).  The information of apparent
Petrosian magnitudes in the five $ugriz$ bands, (Galactic) extinction-
and $K$-corrected absolute magnitudes, and $r$-band Petrosian 90\%
angular radii that include 90\% of Petrosian flux ($\theta_{P90}$)
will be used in the analyses of this work. We take $\theta_{P90}$ from
several size estimations available for SDSS galaxies, since this size
is not sensitive to central concentration (Nair et al. 2010).  The
angular sizes are converted into physical radii as $R_{P90} = d_A(z)
\: \theta_{P90}$, where $d_A$ is the standard angular diameter
distance to redshift $z$.

These galaxies were cross-matched separately with the AKARI/IRC and
AKARI/FIS catalogs, with the matching radius of 5 and 20 arcsec,
respectively. The mean positional errors of sources in these catalogs
are 0.8 and 6 arcsec, and essentially all objects should fall within
the matching radii. [See also Table \ref{tab:akari} for the
instrumental point spread functions (PSF) in each band.]  To avoid
contamination of random associations, we removed any duplications,
i.e., AKARI galaxies having more than two SDSS counterparts, and vice
versa.  We then obtained 390 and 6500 matches of SDSS galaxies with
the IRC and FIS catalogs, respectively.  The expected random
associations are 25.1 and 143.7 for SDSS-IRC and SDSS-FIS, if the
sources are uniformly distributed on the sky.  These numbers are
sufficiently small for studying the overall and statistical trends of
galactic properties.  Because the number of cross-matches with the IRC
catalog is small, we construct the AKARI-SDSS sample based on the
correlation between the SDSS and FIS catalogs, without requiring
detection by IRC.

To keep good quality of the data, we further selected galaxies
satisfying the following conditions: $z \ge 0.005$, $r \le 16.0$ mag,
$\theta_{P90} \ge 6.0''$, and the AKARI/FIS flux quality
flag\footnote{See the release note of the AKARI/FIS catalog for the
  definition of {\it FQUAL}.  It is recommended not to use the flux
  data when {\it FQUAL} $\le 2$ for a reliable scientific analysis.}
{\it FQUAL} $\ge 3$ in at least two FIS bands. The redshift condition
is to remove too close objects whose redshift-based distance is rather
uncertain by the peculiar velocity effect.  The magnitude and size
conditions are to keep good quality of size estimates and to make the
seeing effect negligible.  The condition for {\it FQUAL} is for
reliable SED fittings.  Then 878 galaxies qualified as the final
AKARI-SDSS sample. The numbers of galaxies with {\it FQUAL} $\ge 1$ or
3 in each FIS band are shown in Table \ref{tab:fqual}.  This table
shows that the 90- and 140-$\mu$m bands are more sensitive than the
other two bands and most galaxies are detected in these bands with
{\it FQUAL} $\ge 3$.

To these galaxies we added physical quantities derived from SDSS
spectra by the MPA/JHU group when they are available, by
cross-matching the spectroscopic ID [plate number, Mean Julian Date
(MJD), and fiber number].  Stellar mass ($M_*$, Kauffmann et al. 2003;
Salim et al. 2007), SFR ($\psi$, Brinchmann et al. 2004), and
gas-phase metallicity (Tremonti et al. 2004) of the DR7 version are
available for 550, 567, and 246 galaxies, respectively, in the total
sample of the 878 galaxies.  The metallicity is given in $A_{\rm O}
\equiv $ 12 + log$_{10}$(O/H) in the original data, but in this work
we use metallicity $Z$ converted\footnote{ There are systematic
  uncertainties in estimates of $A_{\rm O}$, $Z$, and their solar
  values (Grevesse et al. 2010; Moustakas et al. 2010).  We will
  mostly discuss correlations between metallicity and other physical
  parameters, and absolute uncertainty does not significantly affect
  our conclusions.\label{fn:metal}} from $A_{\rm O}$ using $A_{\rm O
  \odot} = 8.69$ and $Z_\odot = 0.02$.  We also added the $V$ band
optical depth $\tau_V$ of stellar light attenuation by dust in
galaxies, estimated by Tojeiro et al. (2009) using the VESPA algorithm
(the one-parameter dust model in the RunID=1 catalog), to 534
galaxies.

We also added information of neutral hydrogen (\hins) mass from 21-cm
observations in the literature.  SDSS galaxies were cross-matched with
the optical counterpart positions of 21-cm-detected galaxies in the
data sets of Springob et al. (2005), the ALFALFA survey (Giovanelli et
al. 2007; Kent et al. 2008; Stierwalt et al. 2009), the Equatorial
Survey (Garcia-Appadoo et al. 2009; West et al. 2010), and the GALEX
Arecibo SDSS Survey (Catinella et al. 2010).  We found 257 \hi
flux data available for the AKARI-SDSS sample, and calculated the \hi
mass from the standard formula\footnote{The factor of $(1+z)^{-2}$
  arises as follows. The \hi mass can be written as $M_{\hi} \propto
  d_L^2 \int S_\nu d\nu = d_L^2 \int (d\nu/dV) \: S_\nu \: dV$, where $\nu$
  is the observed frequency. In \hi observations, the velocity $V$ is
  defined as $\nu = \nu_{21} / (1 + V/c)$ where $\nu_{21} \equiv 1.42$
  GHz, and hence $d\nu/dV \propto (1 + V/c)^{-2} \sim (1+z)^{-2}$ when
  the intrinsic velocity dispersion $\Delta V$ in a galaxy is much
  smaller than $cz$.  }:
\begin{eqnarray}
  M_{\hi} = \frac{2.36 \times 10^5}{(1+z)^2} 
  \left(\frac{d_L(z)}{\rm Mpc}\right)^2 
  \left(\frac{S_{\rm int}}{\rm Jy \, km/s}\right) \ M_\odot \ ,
\label{eq:M_HI}
\end{eqnarray}
where $S_{\rm int} \equiv \int S_\nu dV$ is the velocity-integrated
flux density and $d_L$ is the standard luminosity distance.  For
comparison, we also constructed the SDSS-\hi sample, for all SDSS
galaxies with available \hi data without requiring the detection by
AKARI.  Other selection criteria were kept same as the AKARI-SDSS
sample.  The SDSS-\hi sample consists of 2413 galaxies.

\begin{table}
  \caption{The Number of Galaxies against AKARI/FIS Flux Quality Flags}
  \label{tab:fqual}
  \footnotesize
  \begin{center}
    \begin{tabular}{lccccc}
      \hline
      & & \multicolumn{4}{c}{FIS bands [$\mu$m]} \\
      \cline{3-6}
      sample & condition & 65 & 90 & 140 & 160 \\ 
      \hline
      AKARI-SDSS   & {\it FQUAL} $\ge 1$ & 878 & 878 & 878 & 878 \\ 
                   & {\it FQUAL} $\ge 3$ & 178 & 876 & 851 & 136 \\ 
      \hline
      AKARI-HIPASS & {\it FQUAL} $\ge 1$ & 710 & 711 & 711 & 710 \\ 
                   & {\it FQUAL} $\ge 3$ & 173 & 704 & 700 & 239 \\ 
      \hline 
      AKARI-SB     & {\it FQUAL} $\ge 1$ & 24 & 24 & 24 & 24 \\ 
                   & {\it FQUAL} $\ge 3$ & 21 & 22 & 24 & 23 \\ 
      \hline
    \end{tabular}
  \end{center}
\end{table}

\vspace{0.2cm}
\subsubsection{The AKARI-HIPASS Sample}

Since the gas amount of galaxies is crucial for the dust amount, we
also cross-correlated the AKARI galaxies with the HIPASS catalog
(HICAT, Meyer et al. 2004), which is a catalog of $4315$ galaxies
selected by the \hi 21-cm line flux covering the entire southern
sky. We selected HIPASS galaxies having optical counterparts from the
HOPCAT catalog (Doyle et al. 2005), for which the information of 21-cm
line flux, 21-cm redshift, optical magnitudes of $B_j$, $R$, and $I$,
and the major axis size $\theta_{A, \rm HOP}$ of $R$-band images are
extracted from the catalog.  The distances were calculated from 21-cm
redshifts, and then $M_{\hi}$ was calculated from eq. (\ref{eq:M_HI}).
Absolute optical magnitudes were also calculated, correcting the
Galactic extinction but ignoring $K$-correction.

Then the HIPASS galaxies were cross-matched with the AKARI catalogs by
the same procedures as those for the AKARI-SDSS sample.  We found 161
and 1527 matches with the IRC and FIS catalogs, respectively.
Following the AKARI-SDSS sample, we do not require detection by IRC
for the AKARI-HIPASS sample.  To keep good qualities, we further
require the same criteria about redshift ($z \ge 0.005$) and FIS
quality flags ({\it FQUAL} $\ge 3$ at least in two FIS bands) as the
AKARI-SDSS sample, resulting in the final AKARI-HIPASS sample of 711
galaxies.

\begin{figure}
  \begin{center}
    \includegraphics[width=7cm,angle=-90,scale=0.95]{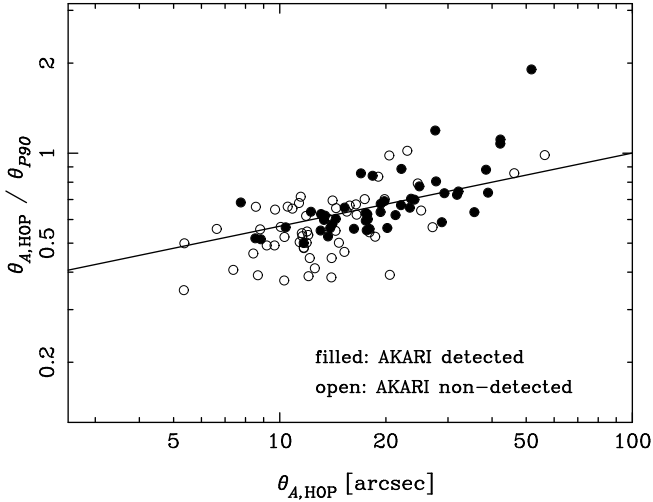}
  \end{center}
  \caption{Correlation between the $r$-band 90\% Petrosian angular
    radius ($\theta_{P90}$) of SDSS and the $R$-band major axis size
    ($\theta_{A, \rm HOP}$) of HOPCAT for galaxies commonly included
    in the SDSS and HIPASS samples. Filled circles are galaxies
    detected by AKARI/FIS.}
  \label{fig:size_comp}
\end{figure}

Because the optical magnitude and size of HOPCAT are from photographic
plates, their quality and reliability should carefully be examined.
To directly compare the AKARI-SDSS and AKARI-HIPASS samples, we
converted the HOPCAT magnitude ($R$) and size ($\theta_{A, \rm HOP}$)
into those in the SDSS catalog ($r$ magnitude and $\theta_{P90}$), and
will use these in the following analyses. For this conversion, we
first searched for galaxies that are common in SDSS and HOPCAT by
optical positions, and found 103 galaxies that satisfy the same
criteria as the AKARI-SDSS galaxies about redshift, magnitude, and
size. In these galaxies 47 are detected by AKARI/FIS.  From comparison
between SDSS $r$ and HOPCAT $R$ bands, we apply the following
conversion: $R({\rm HOP}) - r({\rm SDSS}) = 0.09$, and the dispersion
around this relation is 0.46 mag.  Figure \ref{fig:size_comp} shows
the comparison between the SDSS size $\theta_{P90}$ and the HOPCAT
size $\theta_{A, \rm HOP}$, and the ratio $\theta_{A, \rm HOP} /
\theta_{P90}$ systematically varies with $\theta_{A, \rm HOP}$.  We
therefore convert the HOPCAT size into $\theta_{P90}$ by the following
equation:
\begin{eqnarray}
\log_{10} \left(\frac{\theta_{A, \rm HOP}}{\theta_{P90}}\right) = -0.489 + 
0.245 \ \log_{10} \left( \frac{\theta_{A, \rm HOP}}{\rm arcsec} \right) \ ,
\end{eqnarray}
which is shown in Fig. \ref{fig:size_comp} by the solid line.

\vspace{0.2cm}
\subsubsection{The AKARI-SB Sample}

\begin{figure}
  \begin{center}
    \includegraphics[width=6.5cm,angle=-90,scale=1.0]{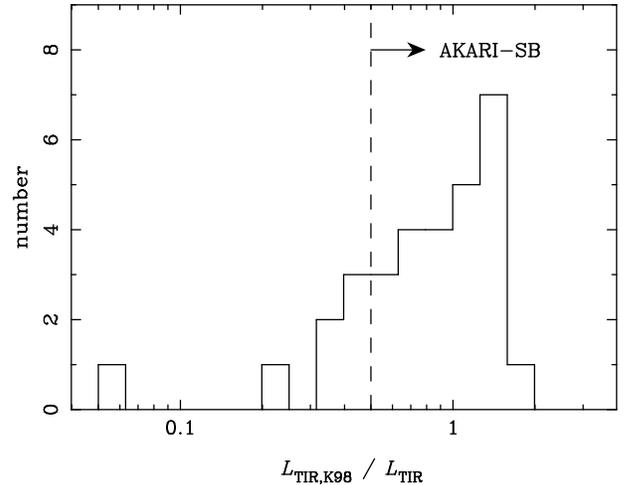}
  \end{center}
  \caption{The histogram of $L_{\rm TIR,K98}/L_{\rm TIR}$ for the K98
    starburst sample, where $L_{\rm TIR, K98}$ is the total infrared
    luminosity from the circumnuclear starburst regions given in 
    Kennicutt (1998) and $L_{\rm TIR}$ is the total infrared
    luminosity from the whole region of galaxies estimated by SED
    fittings to the AKARI data in this work. Only galaxies having
    $L_{\rm TIR,K98}/L_{\rm TIR} \ge 0.5$ are used as the AKARI-SB
    sample.}
  \label{fig:L_TIR_ratio}
\end{figure}

The AKARI/FIS catalog was cross-matched with the 36 galaxies of the
K98 circumnuclear starburst sample by the same procedures as the two
samples described above, and we found that 34 galaxies are detected by
FIS.  To keep the quality of the data, we again adopt the same
condition about the FIS flux quality flags.  For these galaxies we
collected the redshift-independent distance measurements from the NASA
Extragalactic Data base (NED), and adopted the mean value when more
than one measurements are available for a galaxy.  For galaxies that
do not have a redshift-independent distance measurement, we calculated
the distances from redshifts, but NGC 5194 was removed from the sample
because $z < 0.005$.  We will use the information of the angular
diameter size $\theta_D$, SFR surface density $\Sigma_{\rm SFR}$, and
H$_2$ gas mass surface density $\Sigma_{\rm H2}$ given in K98. Here, NGC
3256 was removed because $\theta_D$ is unavailable. 

In K98, SFR and H$_2$ mass were estimated from $L_{\rm TIR}$ and CO
line luminosity, respectively.  As argued in K98, $L_{\rm TIR}$ is a
good SFR indicator for these starburst galaxies because $L_{\rm TIR}$
is dominated by young stars and dust opacity is large.  The
SFR-$L_{\rm TIR}$ conversion factor of K98 assumes the Salpeter (1955)
initial mass function (IMF), and we divided K98 SFRs by 1.5 to match
the SFRs in the AKARI-SDSS sample assuming the Kroupa (2001) IMF
(Brinchmann et al. 2004).  In this case the SFR (denoted as $\psi$) is
related to $L_{\rm TIR}$ as:
\begin{eqnarray}
\frac{L_{\rm TIR}}{L_\odot} = 8.7 \times 10^9
\left(\frac{\psi}{M_\odot \rm \ yr^{-1}}\right) \ .
\label{eq:SFR_L_TIR}
\end{eqnarray}
It should be noted that in this work this relation is used only for
the AKARI-SB sample; we do not use $L_{\rm TIR}$ as a SFR indicator
for the AKARI-SDSS or AKARI-HIPASS samples, because a significant
fraction of $L_{\rm TIR}$ comes from dust heating by relatively aged
stars (see \S \ref{section:L_TIR_L_opt_SFR}). SFRs from SDSS spectra
will be used for the AKARI-SDSS sample.

The surface densities ($\Sigma_{\rm H2}$ and $\Sigma_{\rm SFR}$) given
in K98 are converted into the total quantities ($M_{\rm H2}$ and
$\psi$), e.g., $M_{\rm H2} = \pi R^2 \: \Sigma_{\rm H2}$, where $R =
\theta_D \: d_A /2$ is the physical radius of circumnuclear starburst
regions. Note that the ``total'' quantities in the AKARI-SB sample are
the total within the circumnuclear starburst regions rather than the
entire galaxies.  Since the AKARI photometries are for the whole
region of galaxies, we must examine whether the AKARI fluxes are
mostly coming from the circumnuclear regions.  For this purpose, we
compared the total infrared luminosity from a circumnuclear region,
$L_{\rm TIR, K98}$, of the K98 sample (calculated from $\Sigma_{\rm
  SFR}$ and $\theta_D$) with $L_{\rm TIR}$ for the entire galaxy newly
obtained by our SED fit to the AKARI data (see
\S\ref{section:SED-fit}), in Fig.  \ref{fig:L_TIR_ratio}.  It can be
seen that many galaxies have $L_{\rm TIR} \sim L_{\rm TIR, K98}$,
indicating that the total infrared luminosities of galaxies are
dominated by those from circumnuclear starburst regions. We adopted
the condition of $L_{\rm TIR, K98} \ge 0.5 \ L_{\rm TIR}$ for our
sample, to remove galaxies whose infrared luminosity is mostly emitted
from outer regions. Then finally we obtained the AKARI-SB sample
consisting of 24 galaxies.

\subsection{The Basic Properties of the Sample Galaxies}

\begin{figure*}
  \begin{center}
    \includegraphics[width=13cm,angle=-90,scale=0.5]{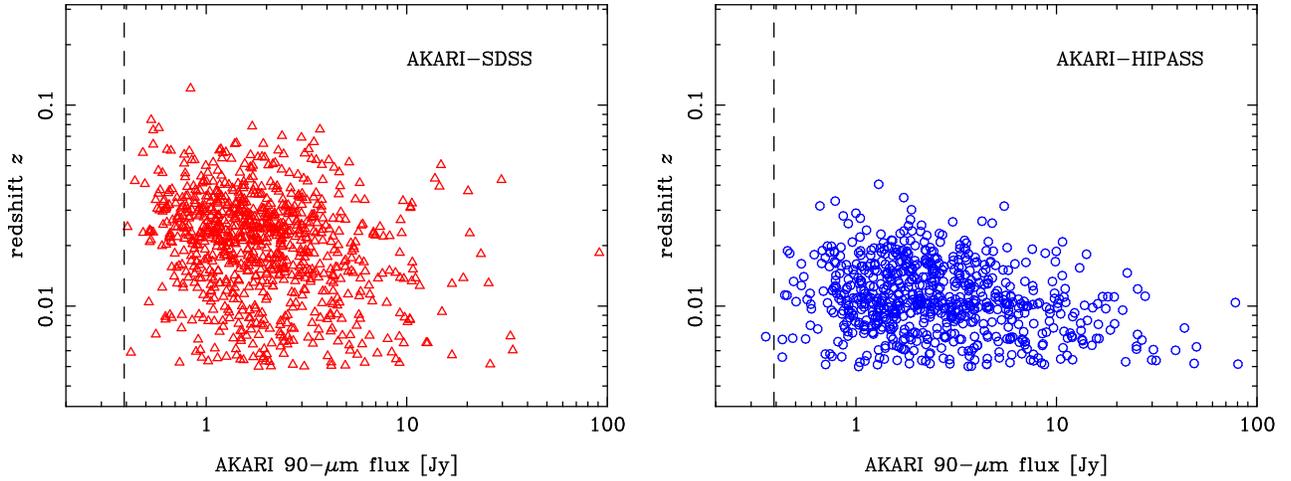}
  \end{center}
  \caption{Redshift versus AKARI 90-$\mu$m band flux for 
the AKARI-SDSS (left) and AKARI-HIPASS (right) samples.
The dashed lines indicate the flux sensitivity limit 
(the flux where the detection completeness is 50\%).}
  \label{fig:F90_z}
\end{figure*}

\begin{figure*}
  \begin{center}
    \includegraphics[width=13cm,angle=-90,scale=0.5]{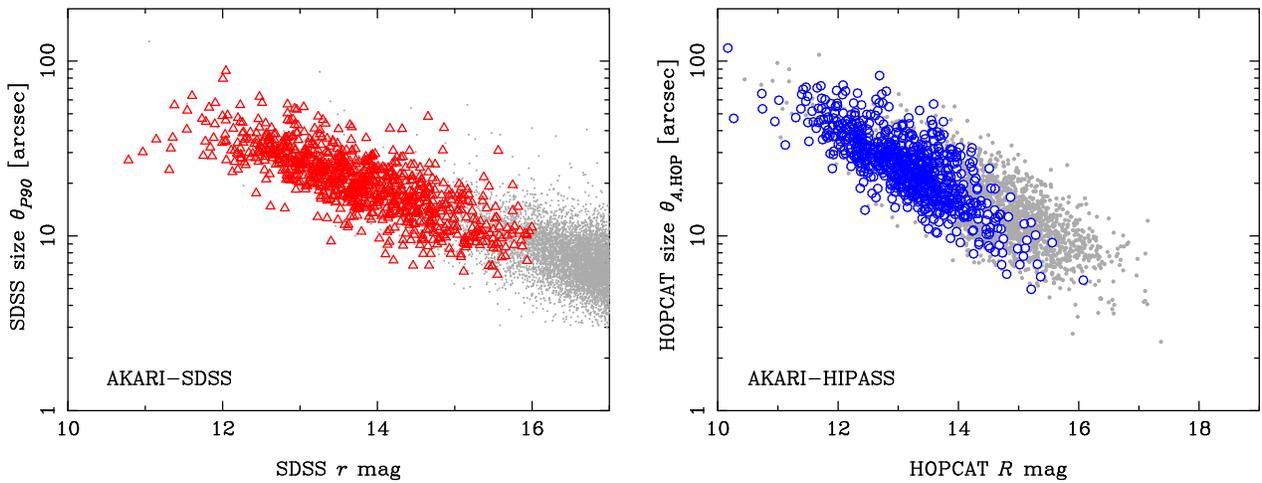}
  \end{center}
  \caption{Angular size versus optical magnitude for the AKARI-SDSS
    (left) and AKARI-HIPASS (right) samples.  The grey dots are
    galaxies in the original SDSS or HIPASS catalog, without requiring
    detection by AKARI.  The grey dots in the left panel are randomly
    selected from the whole SDSS catalog because of the large sample
    size of SDSS.}
  \label{fig:appmag_size}
\end{figure*}

\begin{figure}
  \begin{center}
    \includegraphics[width=7cm,angle=-90,scale=1.0]{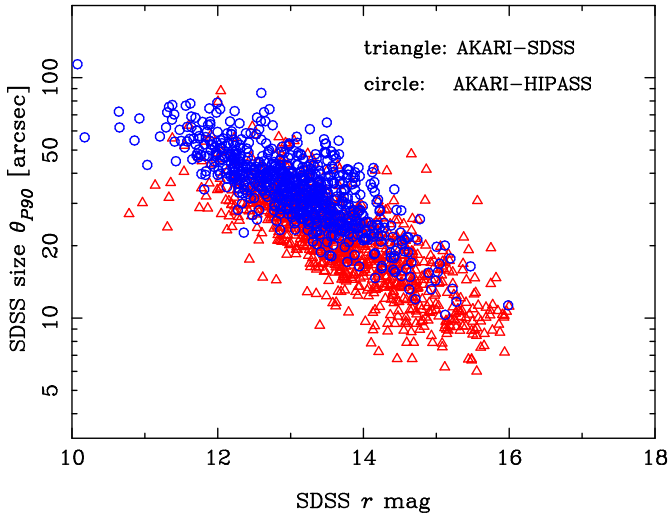}
  \end{center}
  \caption{Comparison between the AKARI-SDSS and AKARI-HIPASS samples
    in the plot of Petrosian 90\% angular radius ($\theta_{P90}$)
    versus SDSS $r$ magnitude, after the conversion of the HOPCAT $R$
    and $\theta_{A, \rm HOP}$ into the SDSS quantities as described in
    the text.}
  \label{fig:appmag_size_comb}
\end{figure}

Figure \ref{fig:F90_z} shows redshift versus the AKARI 90-$\mu$m flux
for the final AKARI-SDSS and AKARI-HIPASS samples. The sensitivity of
the 90-$\mu$m band is the best among the four FIS bands. Sources are
distributed down to the flux sensitivity limit, but the mean flux is
considerably brighter than the flux distribution of all FIS-BSC
sources reported in the release note, because we have set the
condition of {\it FQUAL} $\ge 3$ at least in two FIS bands.  The mean
redshift of the AKARI-SDSS sample ($z \sim 0.03$) is slightly higher
than that of the AKARI-HIPASS sample ($z \sim 0.01$).

Figure \ref{fig:appmag_size} shows angular size versus apparent
optical magnitude ($r$ for SDSS and $R$ for HIPASS).  Compared with
the typical SDSS magnitudes, the AKARI-SDSS galaxies are much
brighter, indicating that the AKARI sensitivity limit is brighter than
SDSS for typical galaxy SEDs.  In Fig. \ref{fig:appmag_size_comb}, we
show the size-magnitude plot of the AKARI-HIPASS galaxies when the $R$
magnitude and $\theta_{A, \rm HOP}$ are converted into $r$ and
$\theta_{P90}$, in comparison with the AKARI-SDSS galaxies. The
distributions of the two samples are similar, indicating that the
conversion of magnitude and size is working well.

\begin{figure*}
  \begin{center}
    \includegraphics[width=13cm,angle=-90,scale=0.5]{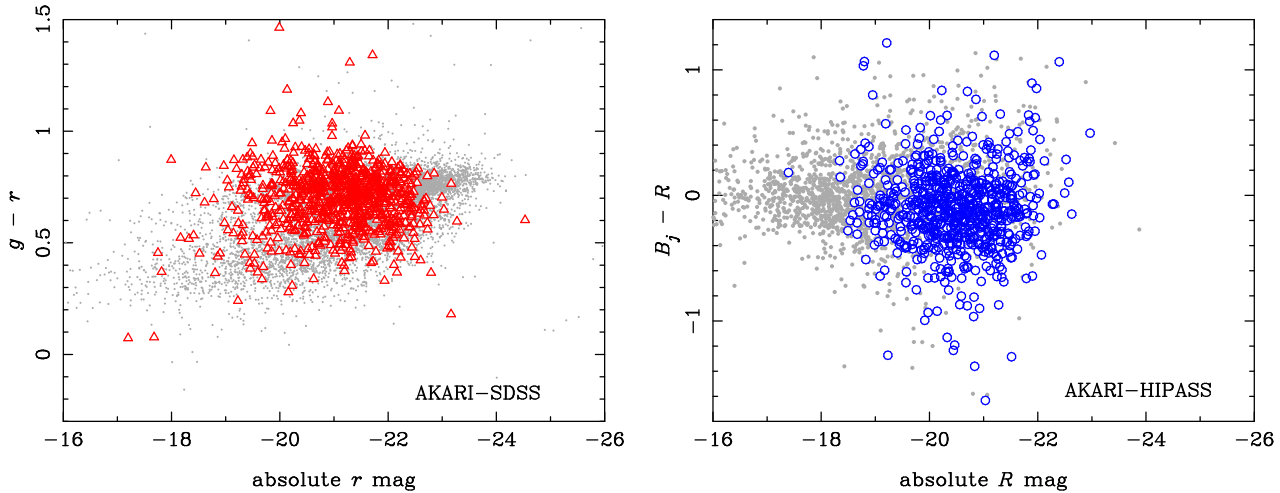}
  \end{center}
  \caption{The same as Fig. \ref{fig:appmag_size}, but for optical
    color versus absolute optical magnitude.  Colors were calculated
    from absolute magnitudes, which have been corrected for the
    Galactic extinction. The $K$-correction has been done for the SDSS
    magnitudes but not for the HIPASS magnitudes.  }
  \label{fig:col_absmag}
\end{figure*}

Figure \ref{fig:col_absmag} shows the optical color versus absolute
magnitude relation (see also Takeuchi et al. 2010).  Our
samples cover a wide range of galaxy colors, though the AKARI-SDSS
galaxies have a slightly redder mean color than the mean SDSS
distribution, indicating that we are sampling dusty galaxies.  Red and
brightest SDSS galaxies are not detected by AKARI, because they are
early-type red-sequence galaxies with no or low star formation
activity.

\begin{figure}
  \begin{center}
    \includegraphics[width=7cm,angle=-90,scale=1.0]{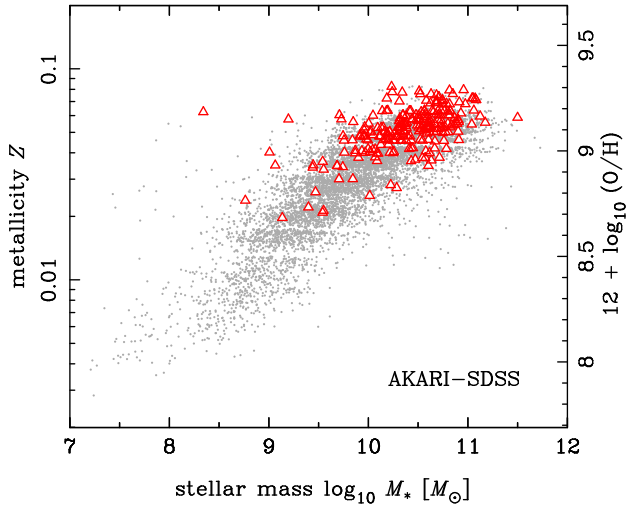}
  \end{center}
  \caption{The stellar mass versus metallicity relation of galaxies in
    the AKARI-SDSS sample. The grey dots are galaxies without
    AKARI-detection, randomly selected from the original SDSS
    catalog.}
  \label{fig:M_Z}
\end{figure}

Figure \ref{fig:M_Z} shows the stellar mass ($M_*$) versus metallicity
relation of the AKARI-SDSS sample, which indicates that we are
sampling relatively massive and high metallicity galaxies compared
with general SDSS galaxies. This is reasonable for our samples
limited by the AKARI sensitivity, because massive star-forming
galaxies are generally dusty and high metallicity is required for a
large amount of dust.

\begin{figure}
  \begin{center}
    \includegraphics[width=7cm,angle=-90,scale=1.0]{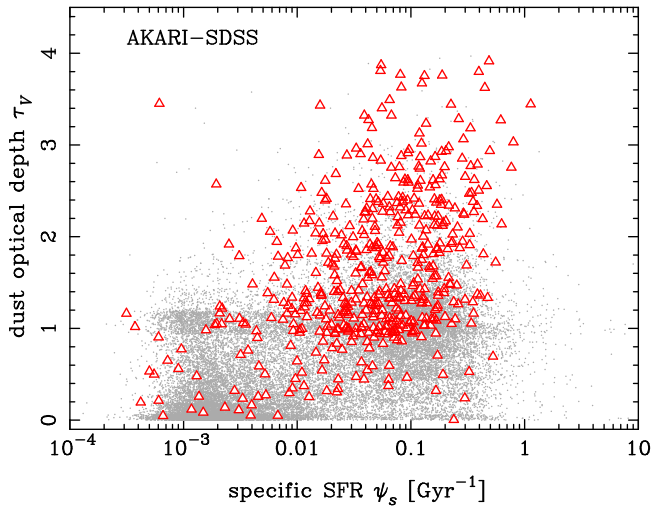}
  \end{center}
  \caption{The same as Fig. \ref{fig:M_Z}, but for $V$-band optical
    depth of dust in galaxies ($\tau_V$) versus specific star
    formation rate ($\psi_s \equiv \psi / M_*$).}
  \label{fig:ssfr_tauV}
\end{figure}

Figure \ref{fig:ssfr_tauV} shows $V$ band attenuation $\tau_V$ versus
specific star formation rate (SSFR) $\psi_s \equiv \psi / M_*$ for the
AKARI-SDSS sample (see also Takeuchi et al. 2010).  The SDSS galaxies
without AKARI detection are bimodally distributed in SSFR reflecting
the well known populations of the red sequence and the blue cloud
(e.g., Strateva et al. 2001; Bell et al. 2004). The AKARI-detected
galaxies are distributed in the high SSFR and high $\tau_V$ region, in
agreement with the general expectation that recent star formation and
a large dust optical depth are necessary for a galaxy to be bright in
far-infrared wavelength.  However, our sample covers a wide range of
SSFR and $\tau_V$, indicating that it includes galaxies having a wide
range of star formation history and dust opacity.

\begin{figure*}
  \begin{center}
    \includegraphics[width=6.7cm,angle=-90,scale=0.95]{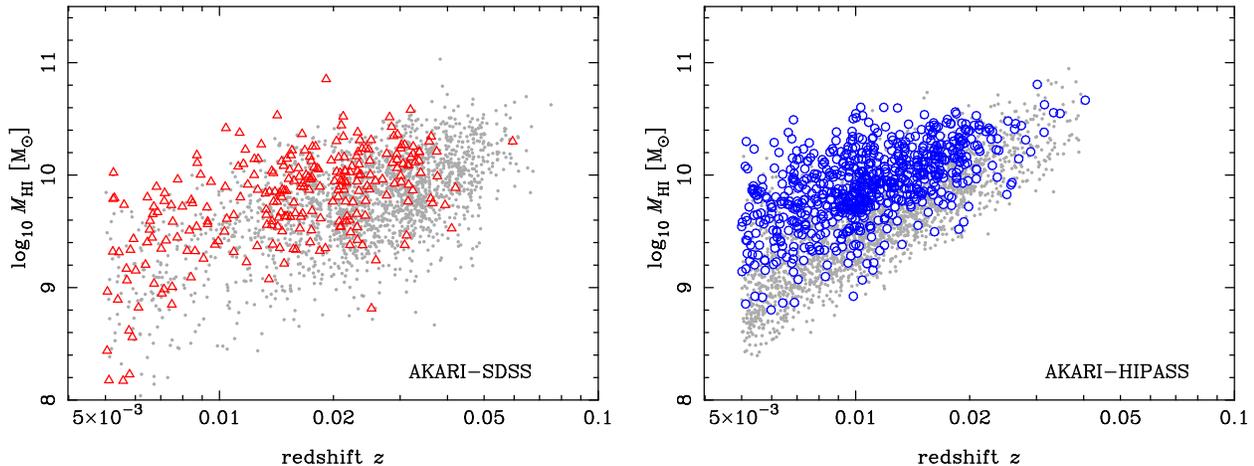}
  \end{center}
  \caption{ Neutral hydrogen mass versus redshift for the AKARI-SDSS
    (left) and AKARI-HIPASS (right) samples.  The grey dots are
    galaxies in the original SDSS or HIPASS catalog, without requiring
    detection by AKARI.}
  \label{fig:z_M_HI}
\end{figure*}

Finally Figure \ref{fig:z_M_HI} shows $M_{\hi}$ versus redshift for the
AKARI-SDSS and AKARI-HIPASS samples. The HIPASS catalog is limited by
\hi flux, and this can clearly be seen in the $z$-$M_{\hi}$ relation
in the right panel. The effect of \hi flux sensitivity limit is also
seen for the AKARI-SDSS sample, though it is less clear because of the
heterogeneous flux limits of the \hi data in the literature.  Galaxies
detected by AKARI have higher \hi mass on average at a fixed redshift
than those without AKARI-detection, and hence the AKARI sensitivity
limit is also important for these samples.

\section{Fitting Physical Dust Models to Infrared SEDs}
\label{section:SED-fit}

\subsection{The Physical Dust Models}

We mainly use the physical dust model of Draine \& Li (2007, hereafter
DL07) to derive physical quantities about dust emission (see also
Weingartner \& Draine 2001; Li \& Draine 2001).  To reduce the number
of model parameters, we choose the $j_M = 5$ model of DL07 for the
Milky Way (MW) that has the PAH mass fraction of $q_{\rm PAH} = 3.19
\%$, because this PAH fraction is close to the median of nearby
galaxies studied by Draine et al. (2007).  The normalization of the
DL07 model is determined by the hydrogen mass $M_{\rm H}$, assuming
the dust-to-hydrogen mass ratio of $f_{d\rm H} \equiv M_d / M_{\rm H}
= 0.01$, but $f_{d\rm H}$ likely depends on properties of galaxies,
especially metallicity.  Therefore in this work we quote the dust mass
as the normalization parameter in the SED fitting, rather than the
hydrogen mass. The dust mass inferred from infrared SED
fittings is denoted by $M_{d\rm IR}$, to make a distinction from the
total dust mass $M_d$ in a galaxy potentially including cold dust that
is not heated by radiation and hence not contributing to infrared
emission.

The basic parameter of infrared emission in the physical dust model
is the heating radiation field strength, $U$, which is the
dimensionless energy density of dust-heating radiation normalized by
the local value of the interstellar radiation field (ISRF) around the
solar neighbourhood.  In reality, the field strength can be different
at different locations in a galaxy, and model infrared emission is
often calculated considering the dust mass distribution $dM_{d\rm
  IR}/dU$ as a function of $U$ (Dale et al. 2001; Dale \& Helou 2002;
DL07).  However, the functional form of $dM_{d\rm IR}/dU$ is highly
uncertain, making it difficult to interpret the results of infrared
SED fittings in a model-independent way.

\begin{figure}
  \begin{center}
    \includegraphics[width=7cm,angle=-90,scale=1.3]{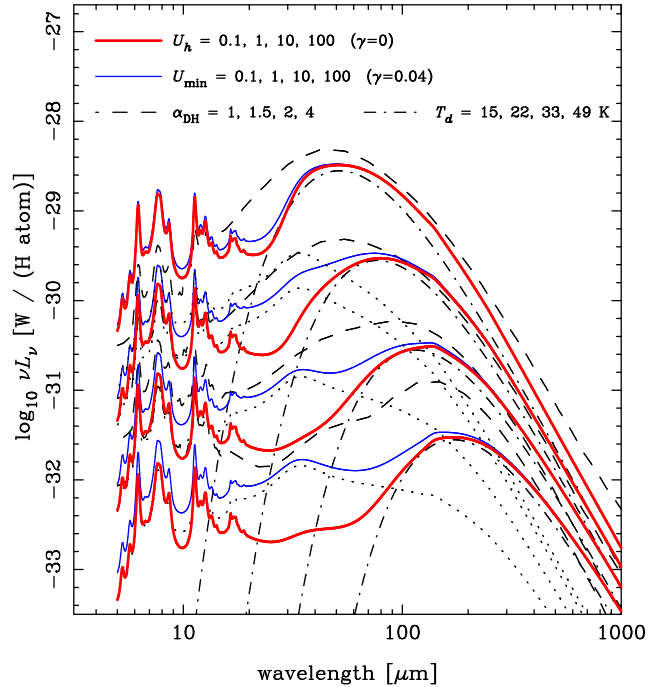}
  \end{center}
  \caption{The infrared SED models of galaxies.  The red thick solid
    curves are the single-$U$ predictions by the physical dust model
    of DL07, for several values of $U_h$ indicated in the figure (from
    bottom to top with increasing $U_h$).  The blue thin curves are
    the same but predictions including the PDR component with $\gamma
    = 0.04$, for several values of $U_{\min}$.  The dotted curves are
    the corresponding PDR components.  The dashed curves are
    predictions by the DH02 model for the four values of $\alpha_{\rm
      DH}$ (from top to bottom with increasing $\alpha_{\rm DH}$).
    The dot-dashed curves are the $\beta = 1.75$ modified blackbody
    spectra for several values of dust temperature $T_d$ (from bottom
    to top with increasing $T_d$).  }
  \label{fig:sed_schem}
\end{figure}

Here we take a different approach by assuming that infrared emission
from a galaxy can be described by a single characteristic heating
radiation strength, $U_h$.  Although it may be difficult to fit
observed infrared SEDs in a wide wavelength range from near- to
far-infrared by a single value of $U$, in this work we concentrate on
the fitting in a relatively narrow range of wavelength covered by
AKARI/FIS (65--160 $\mu$m), corresponding to the broad SED peak of
modified blackbody.  Therefore, $U_h$ in this work should be regarded
as the characteristic value for large grains contributing to the
thermal emission.  An advantage of this approach is that physical
discussion based on the radiation field strength becomes easier. The
SEDs of the single-$U$ DL07 model for several values of $U_h$ are
plotted in Fig. \ref{fig:sed_schem}.

It would be convenient to compare the SEDs of the physical dust model
with the modified blackbody spectra often used in the literature.  It
is known that SEDs around the thermal peak can be approximated by a
modified blackbody\footnote{In the astronomical literature, this is
  also often referred to as ``greybody'', but we avoid this word
  because strictly it is defined as emission with a constant
  emissivity less than one against wavelength.}  with the emissivity
index of $\beta \sim$ 1--2, i.e., flux density per unit frequency
$F_\nu(\nu) \propto Q_{\rm em} (\nu) \: B_\nu(\nu, \: T_d)$, where
$B_\nu$ is the blackbody spectrum of temperature $T_d$ and $Q_{\rm em}
\propto \nu^\beta$ is the emissivity of dust particles (e.g. Dunne et
al. 2000; Kov{\'a}cs et al. 2006).  The total infrared luminosity of
the modified blackbody scales as $L_{\rm TIR} \propto T_d^{4+\beta}$,
and the $\nu F_\nu$ peak scales as $\nu_{\rm peak} \propto T_d$. Since
$L_{\rm TIR} \propto U_h$ for a fixed dust mass, we expect $U_h
\propto \nu_{\rm peak}^{4 + \beta}$. The relation between $U_h$ and
$\nu_{\rm peak}$ of the DL07 model implies $\beta = 1.75$, and for
this value the modified blackbody with $T_d$ = 15, 22, 33, and 49 K
has the same $\nu_{\rm peak}$ as the single-$U$ DL07 model of $U_h =
$ 0.1, 1, 10, and 100, respectively.  These spectra are also plotted
in Fig. \ref{fig:sed_schem}.

\subsection{Fitting to the AKARI/FIS Data}
\label{section:fit-to-FIS}

Because the number of galaxies detected by IRC is small and we
concentrate on $U_h$ for the broad thermal peak in SEDs, we use
only the four AKARI/FIS bands in the SED fits. For a galaxy whose
energy flux per unit frequency is $F_\nu(\nu)$, the $i$-th FIS band
flux quoted in the FIS-BSC catalog (in units of Jy) is calculated as
\begin{eqnarray}
F_i = \frac{1}{\Delta \nu_i} \: \int d\nu \: F_\nu(\nu) \: R_i(\nu) \ ,
\end{eqnarray}
where $\nu$ is the frequency in the observer's frame, $\Delta \nu_i$
is the band width
\begin{eqnarray}
\Delta \nu_i \equiv \nu_c \: \int d\nu \: \nu^{-1} \: R_i(\nu) \ ,
\end{eqnarray}
$\nu_c = c / \lambda_c$ is the central frequency of the band given in
Table \ref{tab:akari}, and $R_i(\nu)$ is RSRF against energy flux for
each AKARI band shown in Fig. \ref{fig:filters}.  The band width is
defined so that $F_i = F_\nu(\nu_c)$ when $\nu F_\nu(\nu)$ is
constant.  The observed energy flux is related to luminosity
$L_\nu(\nu_r; \ M_{d\rm IR}, \: U_h)$ of the DL07 model per unit
rest-frame frequency at a redshift $z$ as
\begin{eqnarray}
  F_\nu(\nu) = \frac{(1+z)}{4 \pi d_L(z)^2} \ 
  L_\nu(\nu_r; \ M_{d\rm IR}, \: U_h) \ ,
\end{eqnarray}
where the rest-frame frequency is $\nu_r = (1+z) \: \nu$.  

Then the best-fit values of $M_{d\rm IR}$ and $U_h$ are calculated by
the standard $\chi^2$ minimization, i.e.,
\begin{eqnarray}
\chi^2(M_{d\rm IR}, U_h) 
= \sum_{i=1}^4 \frac{\left[F_i^{\rm obs} 
- F_i^{\rm model}(M_{d\rm IR}, U_h)\right]^2}{
(\Delta F_{\rm tot, \it i})^2} \ ,
\end{eqnarray}
where the summation is over the four FIS bands. 
When the flux measurement is not reported for a FIS band in the
FIS-BSC catalog, this band was not used in the fitting.  The number
of the FIS bands used for a fitting to a galaxy is denoted by $N_{\rm
  dat}$, and because of our requirement of {\it FQUAL} $\ge 3$ at least in
two of the four FIS bands, $2 \le N_{\rm dat} \le 4$.  There are 3,
44, and 831 galaxies having $N_{\rm dat} = 2$--4 for the AKARI-SDSS
sample and 4, 33, and 674 for the AKARI-HIPASS sample,
respectively. The range of $U_h$ covered in the fittings is $0.1 \le
U_h \le 10^3$, where the lower bound is limited by the available model
library, and the upper bound is large enough because we found no
galaxies having $U_h > 100$.  There are 20 and 24 galaxies whose
best-fit $U_h$ is 0.1 (the lower bound of the model) for the
AKARI-SDSS and AKARI-HIPASS samples, respectively, and the following
statistical analyses are not significantly affected by these small
number of galaxies.  The total infrared luminosity $L_{\rm TIR}$ is
then calculated as the bolometric luminosity of the dust emission from
the best-fit SED model.

The $1\sigma$ error $\Delta U_h$ for $U_h$ was calculated for each
galaxy by the range corresponding to $\Delta \chi^2 = 1$ from the
$\chi^2$ minimum, taking the best-fit normalization (i.e., $M_{d\rm
  IR}$) for each value of $U_h$. The errors of $M_{d \rm IR}$ and
$L_{\rm TIR}$ are correlated with $U_h$, and we estimated them as
follows.  For a fixed value of $U_h$, the error for the normalization
factor can be estimated as (see, e.g., Press et al. 2007) 
\begin{eqnarray}
\frac{\Delta M_{d\rm IR}}{M_{d\rm IR}}
= \frac{\Delta L_{\rm TIR}}{L_{\rm TIR}}
= \frac{1}{M_{d\rm IR}}
\left( \frac{1}{2} \:
\frac{\partial^2 \chi^2}{\partial M_{d\rm IR}^2} \right)^{-\frac{1}{2}} \ .
\end{eqnarray}
In addition, to take into account the error correlated with $U_h$, we
calculated the change of best-fit $M_{d\rm IR}$ and $L_{\rm TIR}$
values when $U_h$ is changed from the best-fit value by $\pm \Delta
U_h$.  Then the quadratic sum of these two types of errors was
calculated as the final error for $M_{d\rm IR}$ and $L_{\rm TIR}$.  We
found that the mean errors of $\log_{10} U_h$ and $\log_{10} M_{d\rm
  IR}$ are $\sim 0.17$ both for the AKARI-SDSS and AKARI-HIPASS
samples, and that for $\log_{10} L_{\rm TIR}$ is smaller than these by
a factor of about four. The error of $L_{\rm TIR}$ is small because it
is directly connected to the observed fluxes, while $M_{d\rm IR}$ and
$U_h$ have similar fractional uncertainties about the SED shape by the
relation $L_{\rm TIR} \propto M_{d\rm IR} \: U_h$.

\begin{figure*}
  \begin{center}
    \includegraphics[width=14cm,angle=-90,scale=1.0]{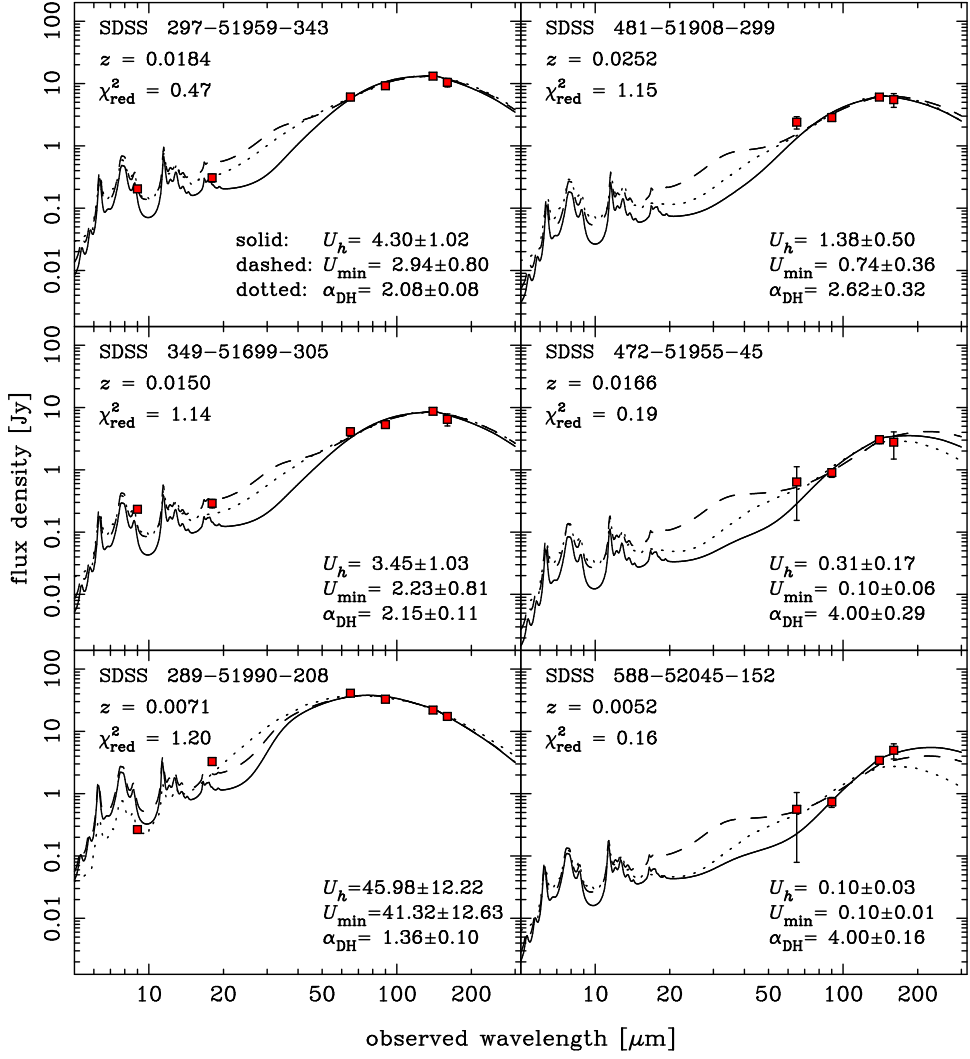}
  \end{center}
  \caption{Examples of the SED fittings to the galaxies in the
    AKARI-SDSS sample. The squares are AKARI data points, and error
    bars are small and difficult to see for most of them.  The three
    different models are used for the fits: the $\gamma=0$ DL07 model
    with the parameter $U_h$ (solid), the $\gamma=0.04$ DL07 model
    with the parameter $U_{\min}$ (dashed), and the DH02 model with
    the parameter $\alpha_{\rm DH}$ (dotted). The best-fit parameters
    and $\chi_{\rm red}^2 = \chi^2/\nu$ (with $\nu = N_{\rm dat}-2$
    degrees of freedom, for the $\gamma=0$ DL07 model) are shown in
    each panel. The unique identification set of plate number, MJD,
    and fiber number of the SDSS spectroscopic database is also shown
    in each panel.  }
  \label{fig:sed_fit}
\end{figure*}

\begin{figure}
  \begin{center}
    \includegraphics[width=7cm,angle=-90,scale=1.36]{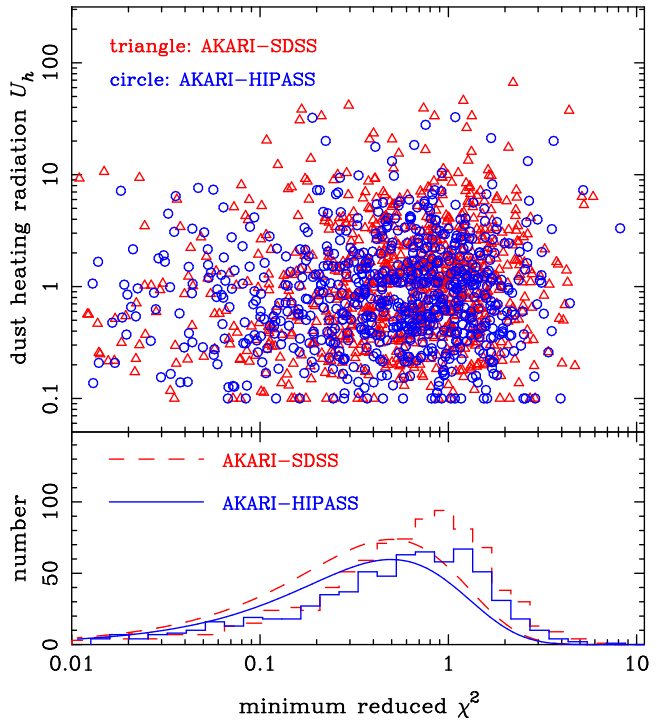}
  \end{center}
  \caption{(Top) Dust-heating radiation field strength ($U_h$) versus
    the minimum reduced chi-square ($\chi^2_{\rm red}$) of SED
    fittings for the AKARI-SDSS and AKARI-HIPASS samples. (Bottom) The
    histogram of the minimum $\chi^2_{\rm red}$.  The curves are the
    theoretically expected distribution from the $\chi^2$ statistics
    for $\nu = (N_{\rm dat} -2)$ degrees of freedom.  Only galaxies
    with $N_{\rm dat} \ge 3$ are shown in this figure.  }
  \label{fig:chi}
\end{figure}

Six examples of the fittings to the AKARI-SDSS galaxies are shown in
Fig. \ref{fig:sed_fit}.  The distributions of $U_h$ and the minimum
reduced chi-square, $\chi^2_{\rm red} \equiv \chi^2/\nu$ for $N_{\rm
  dat} \ge 3$ galaxies are shown in Fig. \ref{fig:chi}, where the
degree of freedom is $\nu = N_{\rm dat} - 2$ for the two-parameter
fittings.  Theoretical expectations of the $\chi^2_{\rm red}$
distribution (weighted sum of those for $N_{\rm dat} = 3$ and 4) are
also shown.  The observed $\chi^2_{\rm red}$ distribution is larger
than the ideal theoretical prediction by a factor of $\sim$ 1.5, but
we consider that this is not unreasonable, if we take into account the
possible systematic errors in the AKARI flux calibration and/or the
possible difference of the single-$U$ DL07 model from the reality.

\subsection{Models versus Observations in the IRC Bands}

\begin{figure}
  \begin{center}
    \includegraphics[width=7cm,angle=-90,scale=1.9]{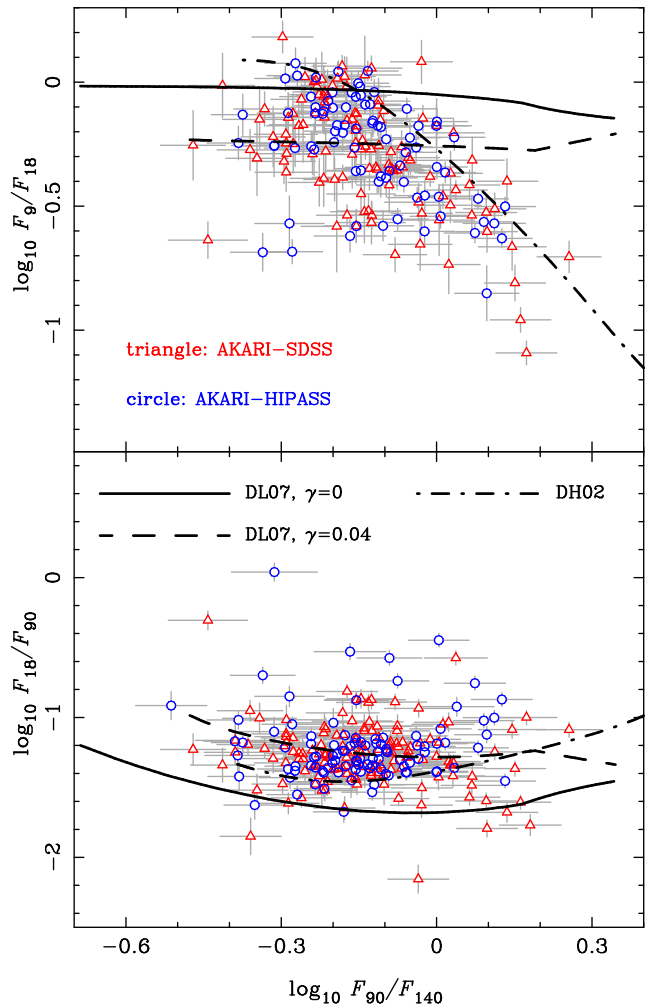}
  \end{center}
  \caption{The color index correlations for the AKARI-SDSS and
    AKARI-HIPASS samples. The flux $F_x$ denotes photometric flux in
    the AKARI $x$-$\mu$m band in units of Jy. The three curves are the
    predicted path along the model parameters by the physical dust
    models.  The solid, dashed, and dot-dashed curves are for the
    $\gamma=0$ DL07 model ($U_h = 0.1$--100), the $\gamma=0.04$ DL07
    model ($U_{\min} = 0.1$--100), and the DH02 model ($\alpha_{\rm
      DH}$ = 0.0625--4), respectively.  The model predictions assume
    $z = 0.02$ that is typical for the observed galaxies.}
  \label{fig:IRC_color}
\end{figure}

In this paper we will concentrate on the SED parameter $U_h$
determined by fits to the AKARI/FIS bands, but it would also be
interesting to compare the observed AKARI/IRC fluxes with those
predicted by the physical dust models.  This is done in
Fig. \ref{fig:IRC_color}, where we show the flux ratios of $F_{18} /
F_{90}$ and $F_9/F_{18}$ against $F_{90}/F_{140}$, where $F_x$ denotes
the AKARI photometric flux of the $x$-$\mu$m band. Here we have used
galaxies with the AKARI/FIS flux quality flag {\it FQUAL} $\ge 3$ both in
the 90- and 140-$\mu$m bands from the AKARI-SDSS and AKARI-HIPASS
samples.  The ratio of $F_{90}/F_{140}$ can be regarded as an
indicator of the dust temperature or $U_h$.  The solid curves are
the path along $U_h$ predicted by the single-$U$ DL07 model, and the
prediction for $F_{18}/F_{90}$ is lower than the mean of the observed
distribution by a factor of about two.  This is reasonable because
mid-infrared emission is likely dominated by dust grains in regions of
stronger heating radiation field, and we must incorporate the
diversity of $U$ to fit SEDs in a wide wavelength range. Here we adopt
two such SED models.

One is the physical dust model by Dale \& Helou (2002, hereafter
DH02), which introduces a power-law distribution of radiation
strength as
\begin{eqnarray}
\frac{dM_{d\rm IR}}{dU} \propto U^{-\alpha_{\rm DH}} \ ,
\end{eqnarray}
and the path along $\alpha_{\rm DH}$ predicted by this model is shown
in Fig. \ref{fig:IRC_color}. The values of $\alpha_{\rm DH} = 4, 3,
1.8$, and 1 roughly correspond to $U_h = 0.6, 1.0, 10$, and 100 of the
single-$U$ DL07 model, respectively, in our SED fits to the four FIS
bands.  This model shows a better agreement with the data than the
single-$U$ DL07 model, demonstrating the necessity of including a
small amount of large $U$ dust component.  On the other hand, the data
show dispersions by about one order of magnitude in both the flux
ratios of $F_9/F_{18}$ and $F_{18}/F_{90}$, indicating the limitation
of one-parameter-family SED models.  The SEDs of the DH02 model are
also shown in Figs.  \ref{fig:sed_schem} and \ref{fig:sed_fit}.

The other model is DL07 using a form of $dM_{d\rm IR}/dU$ proposed by
DL07, which introduces a power-law distribution of $dM_{d\rm IR}/dU$
in the range of $U_{\min} \le U \le U_{\max}$
for a fraction $\gamma$ of the total dust mass, in addition to the
single-$U$ component with $U = U_{\min}$, i.e.,
\begin{eqnarray}
\frac{dM_{d\rm IR}}{dU} &=& (1-\gamma) \: M_{d\rm IR} \:
\delta (U-U_{\min}) 
\nonumber \\ 
&+& \gamma \: M_{d\rm IR} \:
\frac{\alpha-1}{U_{\min}^{1-\alpha} - U_{\max}^{1-\alpha}} \: U^{-\alpha} \ .
\end{eqnarray}
The power-law component is called as the PDR component as it
represents the hot dust in photodissociation regions (PDRs) close to
young massive stars.  We adopt $\alpha = 2$ and $U_{\max} = 10^6$ for
the PDR component following Draine et al. (2007), and now we have
another model parameter $\gamma$ in addition to $U_{\min}$.  (The
model has yet another degree of freedom about the PAH fraction by the
model parameter $j_M$, but we keep $j_M=5$ to reduce the parameter
space to explore.) The PDR component models are available only for
$U_{\min} \le 25$, and we use the $U_{\min} = 25$ template for the PDR
component when $U_{\min} > 25$. We calculated the path along
$U_{\min}$ in the plane of $F_{90}/F_{140}$ versus $F_9/F_{18}$ and
$F_{18}/F_{90}$ for various values of $\gamma$, and found that
$\gamma=0.04$ gives the median of the observed distribution, as shown
in Fig. \ref{fig:IRC_color}.  This value is close to the typical value
found by Drain et al. (2007) for nearby galaxies, though there is a
significant scatter of $\gamma$.  The SED model curves of the
$\gamma=0.04$ model are also shown in Figs.  \ref{fig:sed_schem} and
\ref{fig:sed_fit}.

Although the $\gamma=0.04$ model is better than the single-$U$ model
(i.e., $\gamma = 0$) to reproduce the observed IRC fluxes, we will use
the latter in the analyses below because of the following reasons.
First of all, we want to estimate the characteristic radiation field
strength $U_h$ responsible for the emission around the thermal peak
from large dust grains, and introducing the PDR component in a
model-dependent way may induce some bias in the estimates of $U_h$.
The $\gamma=0.04$ model produces a feature around 30 $\mu$m because of
the rather arbitrary parameter of $U_{\max} = 10^6$, when $U_{\min}$
is small. The best-fit $U_{\min}$ values of the $\gamma = 0.04$ model
tend to be smaller than $U_h$ of the $\gamma=0$ model (see
Fig. \ref{fig:sed_fit}), and we found that a considerable number of
galaxies have the lowest value of $U_{\min} = 0.1$ provided in the
model library. The number of such galaxies can be minimized by using
the $\gamma = 0$ model.  Finally, it should also be noted that the deficit
of the predicted IRC-band flux by the $\gamma=0$ model is at most a
factor of two compared with the mean observed flux.

\section{Theoretical Considerations}
\label{section:theory}

Before we compare the infrared SED parameters with other physical
quantities of galaxies, we describe the general theoretical
expectations.  The basic assumptions here are that dust and heating
radiation sources are uniformly distributed on a disk (but possibly
with different scale heights). This picture is obviously simple, and
allowed only when the infrared SED is determined mostly by the general
ISRF on the global scale in a galaxy, rather than those associated
with small scale structures like individual star forming
regions. However, rather surprisingly, we will find that this simple
picture gives reasonable explanations for most of the data presented
in the paper, and more geometrically complex models are not warranted.

\subsection{Basic Equations of Emission from Dust}

The infrared SED depends on the heating radiation field strength felt
by dust particles, which is defined by the energy density $u_\nu$ or
energy flux $s_\nu$ per unit frequency, and these two are simply
related as $s_\nu = c \: u_\nu$, where $c$ is the speed of light.  The
radiation field can be anisotropic, but anisotropy does not affect the
infrared emission from heated dust, and these quantities are defined
as those integrated over $4\pi$ steradians.  In this work we do not
consider the variation of the SED of the heating radiation field, and
use the same parameter $U_h$ introduced in \S\ref{section:SED-fit},
which is a dimensionless field strength normalized by the local
strength $u_{\nu, 0}$ of the Milky Way ISRF around the solar
neighbourhood, i.e., $u_\nu(\nu) = U_h \: u_{\nu, 0}(\nu)$ and
$s_\nu(\nu) = U_h \: s_{\nu, 0}(\nu)$.

When dust particles of total mass $M_{d\rm IR}$ are exposed to a
heating radiation field strength $U_h$ and emitting in a stationary
state (a good approximation in the AKARI/FIS bands), the bolometric
luminosity $L_{\rm TIR}$ emitted from these particles is given by the
energy balance with the total energy absorbed by dust grains, as
\begin{eqnarray}
  L_{\rm TIR} &=& M_{d\rm IR} \int d\nu \: \kappa_d(\nu) \: s_\nu(\nu) \\
&=& M_{d\rm IR} \: U_h \int d\nu \: \kappa_d(\nu) \: s_{\nu, 0}(\nu) \ ,
\end{eqnarray}
where $\kappa_d$ is the dust mass opacity coefficient (absorption
cross section per unit mass of dust grains, see, e.g., Hildebrand
1983; Dunne et al. 2003).  If we define the frequency-integrated
effective dust mass opacity weighted by $s_{\nu,0}$, as
\begin{eqnarray}
  \kappa_{d, \rm eff} \equiv 
  \frac{ \int d\nu \: \kappa_d(\nu) \: s_{\nu, 0}(\nu)}
  {\int d\nu \: s_{\nu, 0}(\nu)} \ ,
\end{eqnarray}
we obtain
\begin{eqnarray}
  L_{\rm TIR} &=&  \kappa_{d, \rm eff} \: M_{d\rm IR} \: U_h \: s_{\rm bol, 0} \ ,
\label{eq:L_TIR_M_dIR_U_h_s_bol}
\end{eqnarray}
where $s_{\rm bol,0} \equiv \int s_{\nu,0} \: d\nu$ is the bolometric
intensity of the local ISRF, for which we adopt the value of Mathis et
al. (1983): $s_{\rm bol,0} = 0.0259 \ \rm erg \ cm^{-2} s^{-1}$ in
0--13.6 eV (see also Draine et al. 2007).  We can calculate
$\kappa_{d, \rm eff}$ for the physical dust model of DL07 from
eq. (\ref{eq:L_TIR_M_dIR_U_h_s_bol}), which is found to be $\kappa_{d,
  \rm eff} = 1.0 \times 10^4 \ \rm cm^2 \: g^{-1}$. 
In astronomically convenient units, this can also be written as
\begin{eqnarray}
  \frac{L_{\rm TIR}}{L_\odot} = 1.4 \times 10^2 
  \ \frac{M_{d\rm IR}}{M_\odot} \: U_h \ .
\label{eq:L_TIR_M_dIR_U_h}
\end{eqnarray}

Now, adopting the disk geometry, we define the global infrared
radiation field strength of a galaxy by
\begin{eqnarray}
  \Sigma_{\rm TIR} \equiv \frac{1}{2} \: \frac{L_{\rm TIR}}{\pi R^2}
  = \frac{1}{2} \: \kappa_{d, \rm eff} 
  \: \Sigma_{d\rm IR} \: U_h \: s_{\rm bol,0} \ ,
\end{eqnarray}
where $R$ is the characteristic disk radius of a galaxy, and the dust
mass column density of a galactic disk is defined as $\Sigma_{d\rm IR}
\equiv M_{d\rm IR}/(\pi R^2)$. In this work we use the 90\% Petrosian
radius $R_{P90}$ as $R$ for the AKARI-SDSS and AKARI-HIPASS samples,
which includes most of the total luminosity from a galaxy. For the
AKARI-SB galaxies, we use the size of circumnuclear starburst regions
given in K98.  The factor of 1/2 is introduced to approximately
account for the disk geometry where the radiation is emitted into the
two sides of the disk.  The quantity $\Sigma_{\rm TIR}$ has the same
physical dimension as the heating radiation field $s_{\rm bol} = U_h
\: s_{\rm bol,0}$, and we expect some relation between the two if the
infrared emission is mainly determined by the global properties of a
galaxy.  To directly compare with the dimensionless strength $U_h$, we
normalize $\Sigma_{\rm TIR}$ as
\begin{eqnarray}
U_{\rm TIR} \equiv \frac{\Sigma_{\rm TIR}}{s_{\rm bol,0}} 
 = \frac{1}{2} \: \kappa_{d, \rm eff} \: \Sigma_{d\rm IR} \: U_h \ .
\label{eq:U_TIR_Sig_dIR_U_h}
\end{eqnarray}

The effective optical depth $\tau_{d, \rm eff}$ of dust, corresponding
to the light travel distance $l$, is defined as $\tau_{d, \rm eff} =
\kappa_{d, \rm eff} \: \rho_d \: l$, where $\rho_d$ is the spatial
mass density of dust in ISM. The optical depth is related to
dust mass column density and $V$ band extinction as:
\begin{eqnarray}
\tau_{d, \rm eff} = \left(\frac{\rho_d \: l}{\kappa_{d, \rm eff}^{-1}}\right)
= 0.327 \: A_V \ , 
\end{eqnarray}
where $\kappa_{d, \rm eff}^{-1} = 0.48 \ M_\odot \ \rm pc^{-2}$ and we
have used the relation between $A_V$ and hydrogen column density,
$A_V/N_{\rm H} = 5.3 \times 10^{-22} \ \rm mag \; cm^2 \; (H \;
atom)^{-1}$, for the DL07 physical dust model.  We also introduce the
total optical depth $\tautot$ of the disk, which is defined as
\begin{eqnarray}
\tautot \equiv \frac{1}{2} \: \frac{\kappa_{d, \rm eff} \: M_d}{\pi R^2} =
\frac{1}{2} \: \kappa_{d,\rm eff} \: \Sigma_d 
= \frac{\Sigma_d}{\Sigma_{d, \rm crit}}\ ,
\end{eqnarray}
where $M_d$ is the total dust mass in a galaxy, $\Sigma_d \equiv
M_d/(\pi R^2)$, and we have defined the critical dust mass column
density $\Sigma_{d, \rm crit} \equiv 2 \: \kappa_{d, \rm eff}^{-1} =
0.96 \ M_\odot \ \rm pc^{-2}$ for which $\tautot = 1$.  The
factor 1/2 has been introduced because, on average, radiation from
heating sources passes through about half of the total column density
of the dust disk.  It should be noted that here we make a distinction
between $M_{d \rm IR}$ (heated dust mass estimated from infrared
emission) and the total dust amount $M_{d}$ in a galaxy. Generally
$M_{d\rm IR}$ (or $\Sigma_{d\rm IR}$) can be smaller than $M_d$ (or
$\Sigma_d$), if only a fraction of dust particles are radiated by
strong radiation field. The total disk optical depth $\tautot$ is
defined by $M_d$ rather than $M_{d\rm IR}$.  Then we have
\begin{eqnarray}
  U_{\rm TIR} = \tautot \: \frac{\Sigma_{d\rm IR}}{\Sigma_d} \: U_h
  = \tautot \: \frac{M_{d\rm IR}}{M_d} \: U_h  \ .
  \label{eq:U-basic} 
\end{eqnarray}
This is an important equation that will be used to interpret the
observed data in this work.

\begin{figure*}
  \begin{center}
    \includegraphics[width=16.5cm,angle=0]{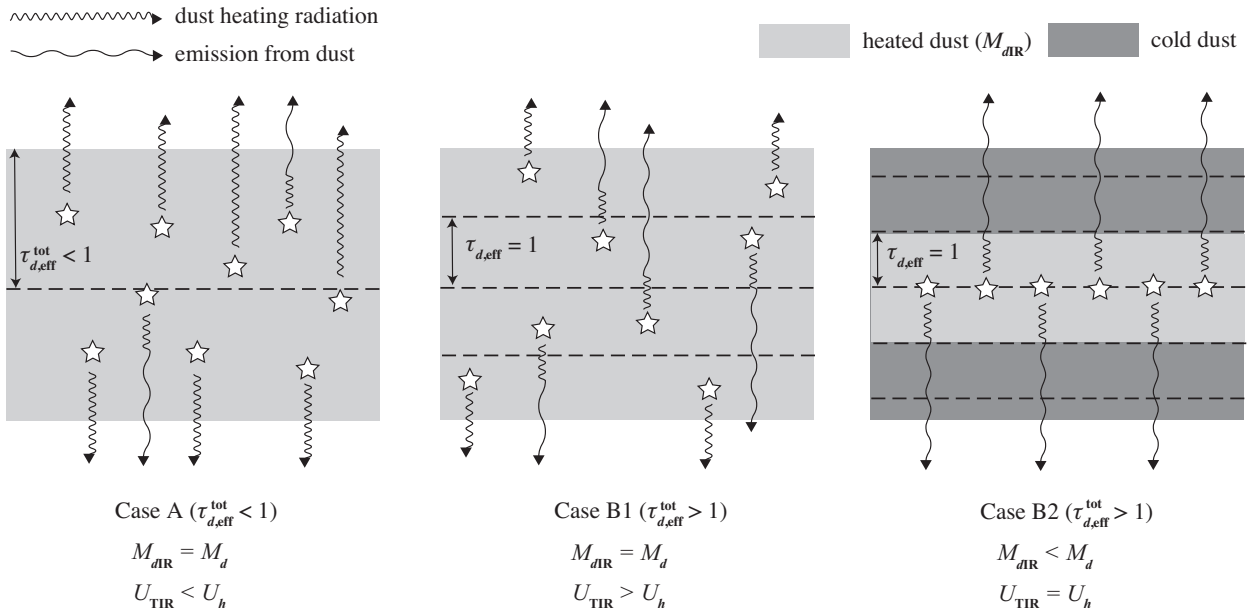}
  \end{center}
  \caption{Schematic pictures for the representative cases of the
    geometrical distributions of dust-heating stars and dust grains,
    in the disk geometry. The normal of the disk plane is along the
    vertical direction.}
  \label{fig:dust_dist_schem}
\end{figure*}

\subsection{Scalings among $U_{\rm TIR}$, $U_h$, $M_{d\rm IR}$,
and $M_d$}
\label{section:scalings}

Eq. (\ref{eq:U-basic}) is a generic equation, but we can consider some
limiting cases with simple scaling laws, corresponding to a particular
geometrical distribution of dust and heating sources.  The schematic
picture to explain these cases is given in
Fig. \ref{fig:dust_dist_schem}.

First we consider the case of small optical depth, $\tautot \ll 1$
(hereafter the case A).  In this case the heating radiation field
strength should be $s_{\rm bol} = U_h \: s_{\rm bol, 0} \sim L_h/(2
\pi R^2)$, where $L_h$ is the bolometric luminosity of heating
radiation from all sources in a galaxy.  This field strength is
relevant for all dust particles in a galaxy, and hence $M_{d\rm
  IR} \sim M_d$, while only a small fraction of the heating radiation
energy is absorbed and converted into the infrared emission, i.e.,
$U_{\rm TIR} \sim \tautot \: U_h < U_h$.

In the opposite limit of large optical depth ($\tautot \gg 1$), most
of heating radiation will be converted into infrared emission, and
hence $L_{\rm TIR} \sim L_h$ and $\Sigma_{\rm TIR} \sim L_h/(2\pi
R^2)$ (hereafter the case B). We can further consider two very
different limiting cases about the relative distributions of dust and
heating sources.  The first case is when dust particles have a similar
disk scale height as the heating sources.  Then all the dust particles
in the disk contribute in the same way to the infrared emission, i.e.,
$M_{d\rm IR} = M_d$.  Then eq. (\ref{eq:U-basic}) tells us $U_{\rm
  TIR} = \tautot \: U_h$, i.e., $U_{\rm TIR} > U_h$, meaning that the
heating radiation strength felt by dust particles ($U_h$) is smaller
than $U_{\rm TIR} = L_h/(2\pi R^2 s_{\rm bol,0})$ by a factor of
$\tautot$. This is because the heating photons are substantially
absorbed after traveling through a distance corresponding to
$\tau_{d,\rm eff} = 1$, and only a fraction ($\sim 1/\tautot$) of
total heating radiation produced in the disk reaches a dust particle.
We call this as the case B1.

Another optically thick case is when the disk scale height of heating
sources is smaller than that of dust particles.  As a simple example,
we assume that the scale height of heating sources is negligibly
small, so that the dust disk behaves like an optically thick screen.
Then the radiation field strength at regions close to the heating
sources should be $s_{\rm bol} = L_h/(2\pi R^2) = \Sigma_{\rm TIR}$,
i.e., $U_h = U_{\rm TIR}$. Then eq. (\ref{eq:U-basic}) requires
\begin{eqnarray}
  \frac{\Sigma_{d\rm IR}}{\Sigma_{d}} &=&
  \frac{M_{d\rm IR}}{M_{d}} = \frac{1}{\tautot} \ ,
\end{eqnarray}
meaning that only a fraction ($\sim 1 / \tautot$) of dust particles
are heated by the radiation, because the radiation field is
significantly damped after traveling through a distance corresponding
to $\tau_{d,\rm eff} = 1$. Infrared emission is mostly from the dust
in the thin layer corresponding to $\tau_{d, \rm eff} \sim 1$ around
the heating sources.  Then it is predicted that the column density of
heated dust is constant at $\Sigma_{d\rm IR} \sim \: \Sigma_{d,\rm
  crit}$, regardless of the total dust column density $\Sigma_d$.  We
call this as the case B2.

Note that $M_{d\rm IR} \sim M_d$ and $U_{\rm TIR} \sim U_h$ can be
realized simultaneously only when $\tautot \sim 1$.

\section{Results and Interpretation}
\label{section:result}

Here we present the main results of this work, i.e., the correlation
between infrared SEDs and various physical properties of galaxies.
The results will be interpreted by the theoretical background
described in the previous section. 

\subsection{Infrared Luminosity versus Optical Luminosity and SFR}
\label{section:L_TIR_L_opt_SFR}

\begin{figure*}
  \begin{center}
    \includegraphics[width=14.5cm,angle=-90,scale=0.5]{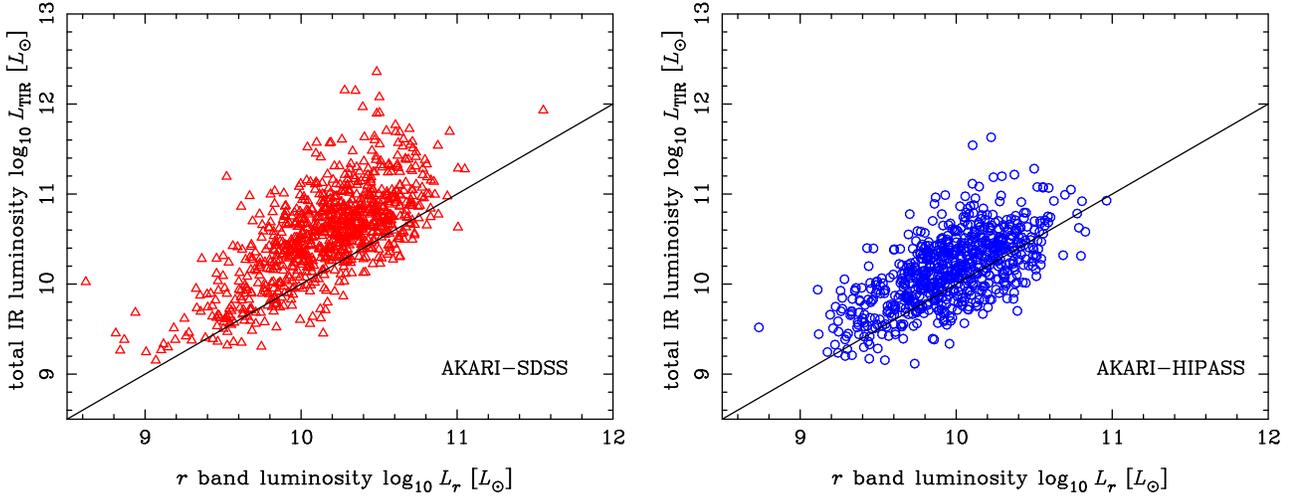}
  \end{center}
  \caption{ Total infrared luminosity ($L_{\rm TIR}$) versus optical
    luminosity $L_r$ ($\nu L_\nu$ at the rest-frame $r$ band) of the
    AKARI-SDSS and AKARI-HIPASS samples.}
  \label{fig:L_opt_L_TIR}
\end{figure*}

First we compare the infrared luminosity $L_{\rm TIR}$ with the
optical luminosity, to examine the dust opacity $\tautot$.  Figure
\ref{fig:L_opt_L_TIR} shows the correlation between $L_{\rm TIR}$ and
the optical $r$-band luminosity ($L_r \equiv \nu L_\nu$ at rest-frame
6231 {\AA}).  We find $L_{\rm TIR} \gtrsim L_r$ for most galaxies,
indicating that energy output by dust emission is similar to, or
larger than that by the direct stellar emission.  It may indicate that
most galaxies have $\tautot \gtrsim 1$, but we can derive this
conclusion only if the $L_{\rm TIR}/L_r$ ratio is determined by the
global properties of galaxies, because $\tautot$ has been defined as
the mean dust column density on a galactic scale. Instead, the $L_{\rm
  TIR}/L_r$ ratio may be determined by the properties of local star
forming regions in a galaxy. We will find that the former
interpretation, $\tautot \gtrsim 1$, is indeed appropriate for the
galaxies studied here.

\begin{figure*}
  \begin{center}
    \includegraphics[width=13.5cm,angle=-90,scale=0.5]{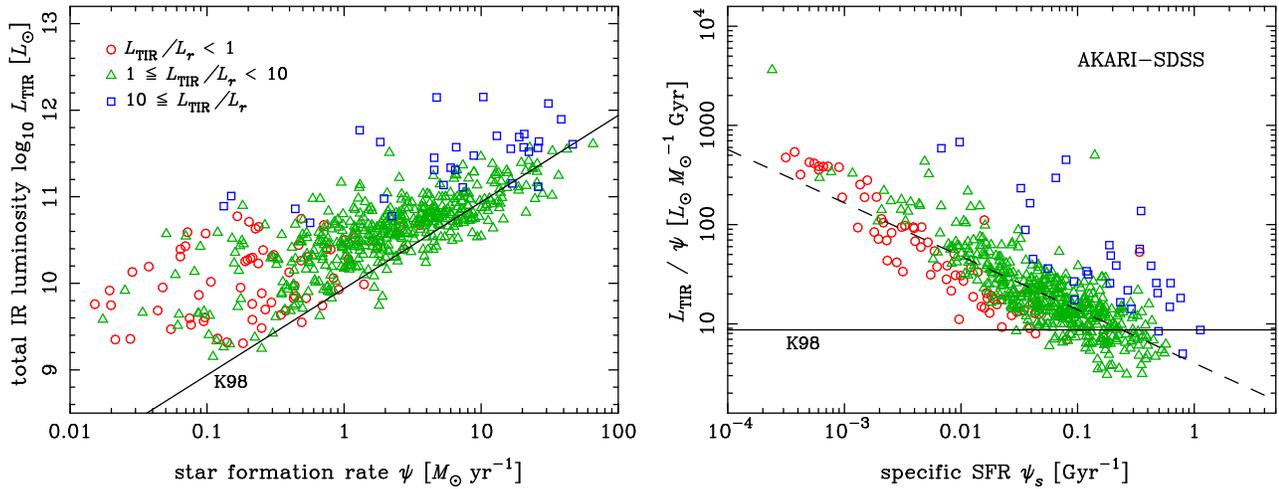}
  \end{center}
  \caption{Correlation between SFR ($\psi$, estimated by SDSS spectra)
    and $L_{\rm TIR}$ (left), and that between SSFR ($\psi_s \equiv
    \psi/M_*$) and $L_{\rm TIR}/\psi$ (right) for the AKARI-SDSS
    sample.  Different symbols are used for different ranges of the
    infrared-to-optical luminosity ratio, $L_{\rm TIR}/L_r$, as shown
    in the left panel. The solid lines are the calibration of $L_{\rm
      TIR}/\psi$ by Kennicutt (1998, but corrected for the Kroupa IMF)
    for $L_{\rm TIR}$ as an SFR indicator.  The dashed line in the
    right panel is the best-fit power-law to the galaxies with $L_{\rm
      TIR}/L_r < 10$.  }
  \label{fig:SFR_L_TIR}
\end{figure*}

Next we examine the correlation between $L_{\rm TIR}$ and SFR ($\psi$,
estimated from SDSS spectra) of the AKARI-SDSS sample, shown in the
left panel of Figure \ref{fig:SFR_L_TIR}, to examine the origin of
dust heating radiation.  The total infrared luminosity $L_{\rm TIR}$
is often used as a SFR indicator by eq. (\ref{eq:SFR_L_TIR}), and its
validity and limitation have been discussed in the literature (e.g.,
Takeuchi et al. 2005, Salim et al. 2009; Boquien et al. 2010; Goto et
al. 2011b).  It cannot be a good SFR indicator when star formation
activity is low and dust heating radiation is significantly
contributed from relatively aged stars. This trend is clearly seen in
this figure; $L_{\rm TIR}/\psi$ increases with decreasing SFR.  The
solid line is the $L_{\rm TIR}/\psi$ ratio of
eq. (\ref{eq:SFR_L_TIR}), which is the value expected when the dust
heating radiation is dominated by young stellar populations with ages
of 10--100 Myr (Kennicutt 1998, converted to the Kroupa IMF
here). This $L_{\rm TIR}/\psi$ ratio is observed only for high SFR
galaxies. Different symbols are used in this figure for different
ranges of $L_{\rm TIR}/L_r$, and the trend of larger $L_{\rm TIR}/L_r$
for larger SFR can be seen, in agreement with the well-known fact that
actively star-forming galaxies are generally dusty.

Since SSFR ($\psi_s \equiv \psi / M_*$) is a good measure of relative
abundances of young and old stellar populations, we expect that
$L_{\rm TIR}/\psi$ is tightly correlated with SSFR. This is indeed
seen in the right panel of Figure \ref{fig:SFR_L_TIR}, especially for
galaxies having $L_{\rm TIR}/L_r < 10$.  On the other hand, dusty
galaxies having $L_{\rm TIR}/L_r > 10$ deviate from the
correlation showing particularly high $L_{\rm TIR}$, and this is most
likely explained by the hidden activity of star formation that cannot
be captured by optical emission because of large extinction. The tight
correlation of galaxies with $L_{\rm TIR}/L_r < 10$ can be fit as
\begin{eqnarray}
\frac{L_{\rm TIR}}{\psi} = 4.0
\left( \frac{\psi_s}{\rm Gyr^{-1}} \right)^{-0.54} \
L_\odot \ M_\odot^{-1} \ \rm Gyr \ .
\end{eqnarray}

This result indicates that SSFR can be used as an indicator of
contribution from young stellar population to dust-heating
radiation. The infrared luminosity $L_{\rm TIR}$ can be used as a
simple SFR indicator only for actively star-forming galaxies of
$\psi_s \gtrsim 0.1 \ \rm Gyr^{-1}$, while for other galaxies relatively
aged stars significantly contribute to the dust-heating radiation.
The tight $L_{\rm TIR}/\psi$ versus $\psi_s$ correlation may be useful
for parameter estimations for observed galaxies, because if two of the
three parameters ($M_*$, SFR, and $L_{\rm TIR}$) are known, the
remaining parameter can be deduced by solving the $L_{\rm TIR}/\psi$ -
$\psi_s$ relation.  For example, if one knows $L_{\rm TIR}$ and $M_*$
from observations, SFR can be solved from this relation, which should
be a better estimation than simply using eq. (\ref{eq:SFR_L_TIR}).

\subsection{Infrared SED versus  Infrared Luminosity and SSFR}
\label{section:L_TIR_U_h}

\begin{figure*}
  \begin{center}
    \includegraphics[width=14.5cm,angle=-90,scale=0.5]{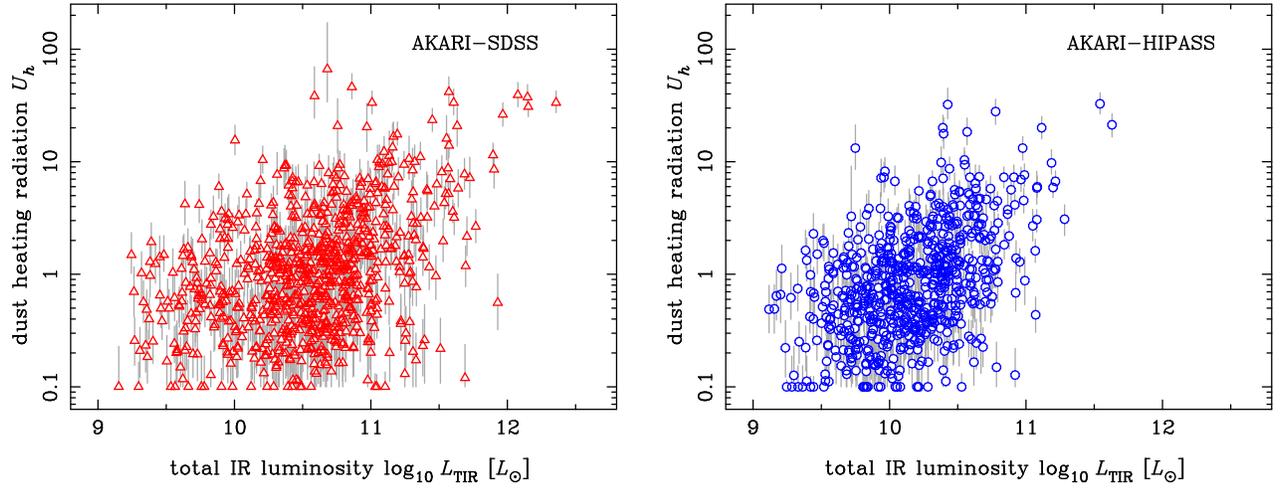}
  \end{center}
  \caption{ Dust-heating radiation field strength ($U_h$, estimated by
    infrared SED fittings) versus total infrared luminosity $L_{\rm
      TIR}$, for the AKARI-SDSS (left) and AKARI-HIPASS (right)
    samples. }
  \label{fig:L_TIR_U_h}
\end{figure*}

Figure \ref{fig:L_TIR_U_h} shows the correlation between $L_{\rm TIR}$
and $U_h$, for the AKARI-SDSS and AKARI-HIPASS samples.  As mentioned
in \S\ref{section:intro}, the correlation between the two (larger
$U_h$ for larger $L_{\rm TIR}$) is widely known, and the trend is also
confirmed for both the two samples studied here. However, there is a
considerable scatter and it is difficult to predict $U_h$ with a good
accuracy only by the information of $L_{\rm TIR}$. Furthermore, we
find a significant offset of the mean relation between the two
sample. Since the AKARI-SDSS galaxies are more distant, their mean
$L_{\rm TIR}$ is larger than that of the AKARI-HIPASS galaxies, but the
distribution of $U_h$ is not significantly different. This indicates
that the $L_{\rm TIR}$-$U_h$ or $L_{\rm TIR}$-$T_d$ relation is
sensitive to the sample selections.

\begin{figure*}
  \begin{center}
    \includegraphics[width=14.5cm,angle=-90,scale=0.5]
        {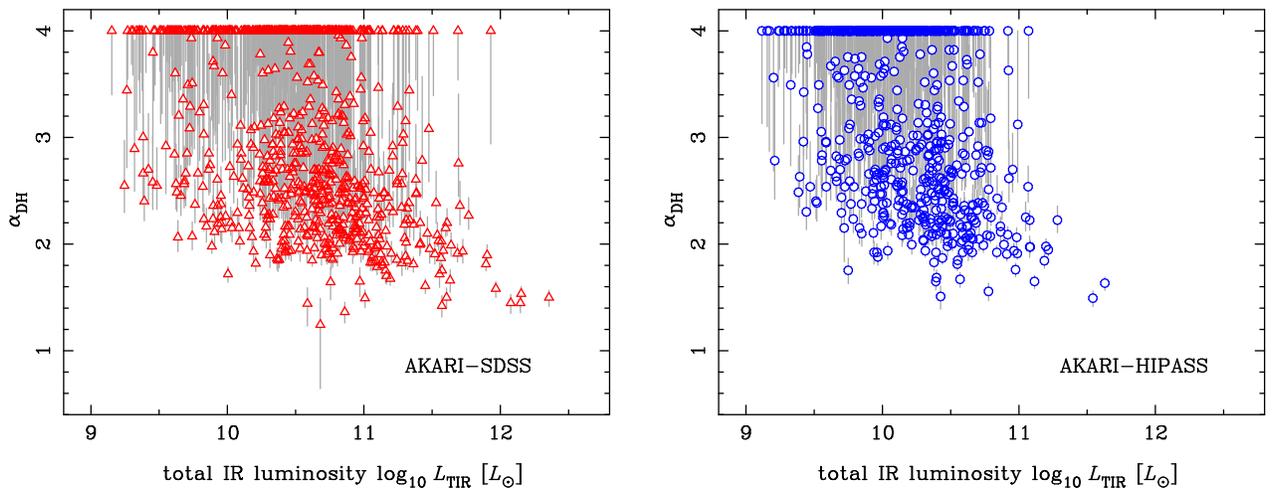}
  \end{center}
  \caption{The same as Fig. \ref{fig:L_TIR_U_h}, but for the
    correlation of $\alpha_{\rm DH}$ (obtained by the SED fittings
    with the DH02 model) versus $L_{\rm TIR}$. }
  \label{fig:L_TIR_alpha_DH}
\end{figure*}

Figure \ref{fig:L_TIR_alpha_DH} shows the correlation between $L_{\rm
  TIR}$ and $\alpha_{\rm DH}$ of the DH02 model fits.  It is
interesting to compare the $\alpha_{\rm DH}$ distribution of our
samples with that of the IRAS $F_{60} > $ 1.2 Jy sample (Fisher et
al. 1995).  Converting the distribution of $F_{60}/F_{100}$ flux ratio
of this sample given in Chapman et al. (2003) into $\alpha_{\rm DH}$
using the relation in Table 2 of Dale et al. (2001)\footnote{ Here we
  ignored the $K$-correction, which is a good approximation for the
  low-redshift sample.}, the mean is $\alpha_{\rm DH} \sim 2$ and
there are few galaxies having $\alpha_{\rm DH} > 2.5$ in this
IRAS-based sample. This is in contrast to the AKARI-SDSS and
AKARI-HIPASS samples that include many galaxies at $\alpha_{\rm DH} >
2.5$. This is likely because the IRAS sample was selected at the IRAS
60-$\mu$m band, while many of the galaxies in our sample are detected
in the AKARI 90- and 140-$\mu$m bands with the FIS flux quality flag
{\it FQUAL} $\ge$ 3 (see Table \ref{tab:fqual}).  Therefore our
samples include galaxies having colder dust temperatures than the
IRAS-based sample.  Note that the errors of $\alpha_{\rm DH}$
(calculated in the same way as $U_h$) are large when $\alpha_{\rm DH}
\gtrsim 3$, because the dependence on $\alpha_{\rm DH}$ of the DH02
templates becomes weaker.

\begin{figure}
  \begin{center}
    \includegraphics[width=7cm,angle=-90,scale=1.0]{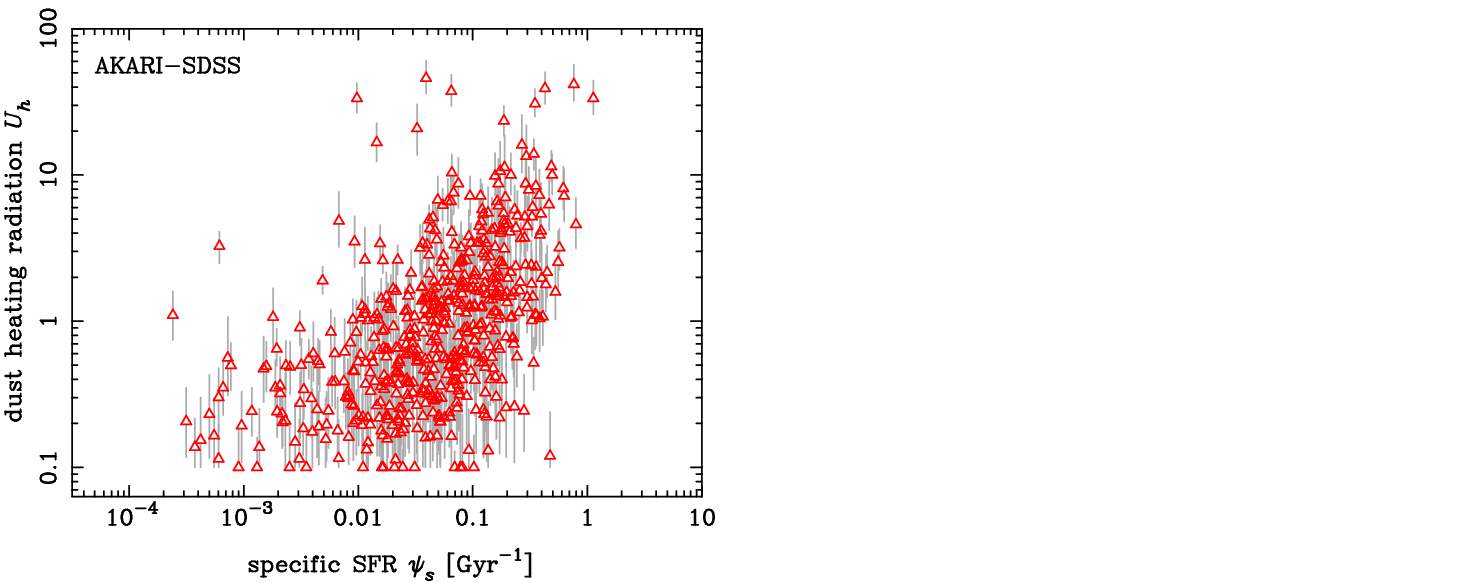}
  \end{center}
  \caption{ Dust-heating radiation strength ($U_h$, estimated by
    infrared SED fittings) versus specific star formation rate
    ($\psi_s \equiv \psi / M_*$, $\psi$ and $M_*$ from SDSS spectra)
    for the AKARI-SDSS sample. }
  \label{fig:ssfr_U_h}
\end{figure}

Figure \ref{fig:ssfr_U_h} shows the correlation between SSFR and $U_h$
for the AKARI-SDSS sample, using SDSS SFR and stellar mass.  This
correlation is physically more reasonable than $L_{\rm TIR}$-$U_h$,
because both SSFR and $U_h$ are intensive physical quantities (i.e.,
not depending on the system size or the amount of material). In fact,
we see a better correlation in Fig. \ref{fig:ssfr_U_h} than the
$L_{\rm TIR}$-$U_h$ correlation: larger $U_h$ for galaxies having
larger SSFR.  However the slope of the correlation becomes steeper at
$\psi_s \gtrsim 0.1 \ \rm Gyr^{-1}$, and it is still difficult to
predict $U_h$ reliably simply from SSFR. It should also be noted that
SSFR is not a directly observable quantity, and physical relation
between SSFR and $U_h$ is not clear. SSFR is related to star formation
history in a galaxy and hence it would affect the SED of dust-heating
radiation field. However, a key quantity controlling the radiation
field strength is the geometrical size, which is not directly related
with SSFR.  Correlations between $U_h$ and more directly observable
quantities based on better physical connections are worth to seek for.

\subsection{Infrared SED versus Global Infrared Radiation Field Strength}
\label{section:U_TIR_U_h}

\begin{figure*}
  \begin{center}
    \includegraphics[width=14.5cm,angle=-90,scale=0.5]{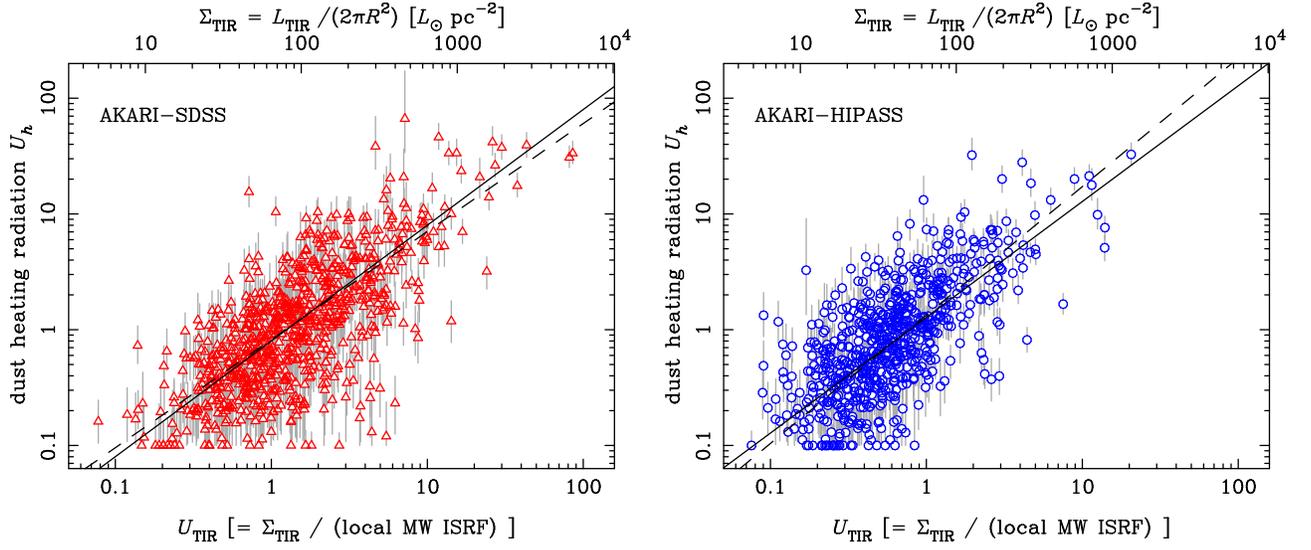}
  \end{center}
  \caption{ Dust-heating radiation field strength ($U_h$, estimated by
    infrared SED fittings) versus global infrared radiation field
    strength [$\Sigma_{\rm TIR} = L_{\rm TIR}/(2\pi R^2)$], for the
    AKARI-SDSS (left) and AKARI-HIPASS (right) samples.  The parameter
    $U_{\rm TIR}$ is $\Sigma_{\rm TIR}$ normalized by the bolometric
    strength of the local (solar neighborhood) ISRF in the Milky Way,
    which can directly be compared with $U_h$.  The solid and dashed
    lines are the best-fit linear ($U_h \propto U_{\rm TIR}$) and
    power-law ($U_h \propto U_{\rm TIR}^b$) relations, respectively.}
  \label{fig:U_TIR_U_h}
\end{figure*}

Figure \ref{fig:U_TIR_U_h} shows the correlation between $U_h$ and
$U_{\rm TIR}$, i.e., the galactic-scale infrared radiation field
strength $\Sigma_{\rm TIR} \equiv L_{\rm TIR}/(2\pi R^2)$ normalized
by the solar neighbourhood value of ISRF.  It can be seen that these
two parameters are correlated much better than the $L_{\rm TIR}$-$U_h$
or $\psi_s$-$U_h$ relations.  The relation is well described by the
linear relation of $U_h \propto U_{\rm TIR}$, with the ratio
$U_h/U_{\rm TIR} \sim 1$.  This result is in nice agreement with the
prediction in the case B2 discussed in \S\ref{section:scalings}, and
it strongly indicates that (1) integrated infrared SED of a galaxy is
mostly determined by the general ISRF on the global scale of a galaxy,
and (2) dust-heating sources are deeply embedded in an optically-thick
dust disk, i.e., the scale height of heating sources being smaller
than one optical depth of the dust disk.

There are several independent observational evidences supporting this
picture.  Walterbos \& Greenawalt (1996) and Sun \& Hirashita (2010)
argued for the dominant role of the general ISRF in the determination
of infrared SEDs. The geometry of the case B2 is consistent with the
smaller scale height of molecular gas than that of H\emissiontype{I}
in the Galaxy (Malhotra 1994, 1995; Binney \& Merrifield 1998),
considering that young stars are generally born in dense molecular
gas.  Here we implicitly assumed that dust-to-gas ratio is fairly
constant throughout the Galaxy, which is reasonable as inferred from
the well-known good correlation between total (i.e., \hi plus H$_2$)
hydrogen column density and color excess of extinction (Bohlin et
al. 1978; Heithausen \& Mebold 1989; Rachford et al. 2009).  Finally
we note that the original form of the Schmidt law ($\psi \propto
\rho_{\rm gas}^2$ where $\rho_{\rm gas}$ is spatial gas density) was
proposed to explain the smaller vertical scale height of young stars
than that of interstellar gas (Schmidt 1959; Fuchs et al. 2009).

\begin{figure*}
  \begin{center}
    \includegraphics[width=14.5cm,angle=-90,scale=0.5]{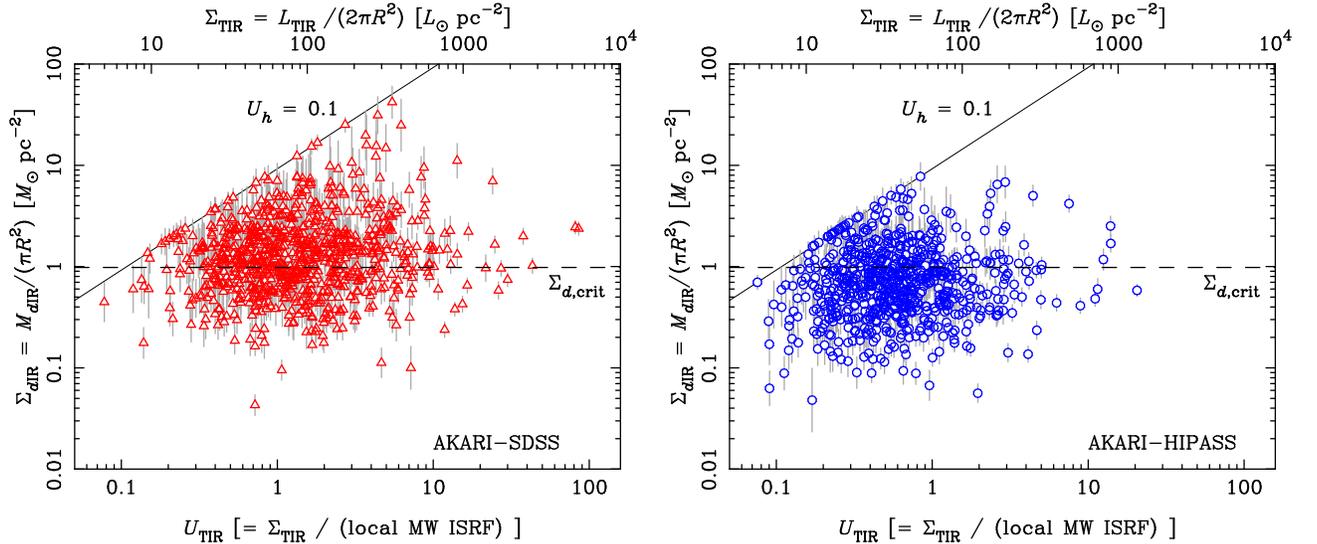}
  \end{center}
  \caption{ Column density of heated dust $\Sigma_{d\rm IR}$ versus
    global infrared radiation field strength ($U_{\rm TIR}$ or
    $\Sigma_{\rm TIR}$ for the bottom or top ordinate), for the
    AKARI-SDSS (left) and AKARI-HIPASS (right) samples.  The critical
    dust column density, $\Sigma_{d, \rm crit} = 0.98 \: M_\odot \rm
    pc^{-2}$ corresponding to the effective dust optical depth
    $\tautot = 1$, is indicated by the dashed line.  The solid line
    shows the locations when $U_h = 0.1$, and there is no galaxies in
    the upper-left region because the physical dust model is limited
    to $U_h \ge 0.1$. The errors are from those for $M_{d\rm IR}$.}
  \label{fig:U_TIR_Sig_Md}
\end{figure*}

In the case B2, we expect that the dust column density inferred from
infrared SED fits, $\Sigma_{d\rm IR}$, is constant around $\Sigma_{d,
  \rm crit} \equiv 2 \: \kappa_{d, \rm eff}^{-1} = 0.96 \ [M_\odot \: \rm
pc^{-2}]$. This is checked by Fig. \ref{fig:U_TIR_Sig_Md}, where
correlation between $U_{\rm TIR}$ and $\Sigma_{d\rm IR}$ is shown. As
expected, the values of $\Sigma_{d\rm IR}$ do not show any trend
against $U_{\rm TIR}$, with the mean value close to $\Sigma_{d, \rm
  crit}$. Many of galaxies having large deviations from $\Sigma_{d,
  \rm crit}$ have large observational errors.  Note that no galaxies
are found in the region of $U_h < 0.1$ simply because the fit is
limited by the minimum value $U_h = 0.1$ of the physical dust model.

We fit the $U_{\rm TIR}$-$U_h$ relation by a linear ($U_h = 10^a
U_{\rm TIR}$) relation and the results are presented in Table
\ref{tab:fits} for the AKARI-SDSS and AKARI-HIPASS samples.  Here we
considered only the errors of $\log_{10} U_h$, because the errors of
$\log_{10} L_{\rm TIR}$ are much smaller, as mentioned in \S
\ref{section:fit-to-FIS}.  The proportionality constant is different
by just a factor of 1.5 for the two samples, indicating that our
result is not seriously affected by the selection bias.  We also tried
a power-law ($U_h = 10^a U_{\rm TIR}^b$) relation and a
three-dimensional plane fit in the space of $\log_{10} R$, $\log_{10}
L_{\rm TIR}$, and $\log_{10} U_h$.  The results are shown in Table
\ref{tab:fits}, and these fits also indicate that the simple relation
of $U_h \propto U_{\rm TIR} \propto L_{\rm TIR}/R^2$ is a good
description of the observed correlation.

We estimated the physical dispersion of $U_h$ from the mean relation
as follows.  We define $\Delta_{\lg U}^{\rm obs}$ as the observed
deviation of $\log_{10} U_h$ of each galaxy from the best-fit
relations.  Then the observed median of $|\Delta_{\lg U}^{\rm obs}|$
was converted into the observed standard deviation $\sigma_{\lg
  U}^{\rm obs}$ by using the Gaussian relation (median of $|x|$ is
$0.674 \: \sigma$).  The use of the median is to avoid the effect of a
small number of outliers having large $|\Delta_{\lg U}^{\rm obs}|$.
The quantity $\sigma_{\lg U}^{\rm obs}$ includes the observational
errors of $U_h$ in the SED fittings, and the intrinsic (i.e.,
physical) standard deviation $\sigma_{\lg U}^{\rm int}$ was calculated
as
\begin{eqnarray}
\sigma_{\lg U}^{\rm int} = \left[ \: (\sigma_{\lg U}^{\rm obs})^2 
  - \langle \sigma_{\lg U}^{\rm fit} \rangle^2 \: \right]^{\frac{1}{2}} \ ,
\end{eqnarray}
where $\langle \sigma_{\lg U}^{\rm fit} \rangle$ is the mean of
1$\sigma$ errors of $\log_{10} U_h$ in the SED fits.  The results for
$\sigma_{\lg U}^{\rm obs}$, $\sigma_{\lg U}^{\rm int}$ and $\langle
\sigma_{\lg U}^{\rm fit} \rangle$ are shown in Table \ref{tab:fits},
and $\sigma_{\lg U}^{\rm int} \sim 0.3$ is indicated for all cases.
This corresponds to $\sim$13 \% dispersion in terms of the modified
blackbody temperature, using $U_h \propto T_d^{4+\beta}$ and $\beta =
1.75$.

\begin{table}
  \caption{Functional Fits to the $U_{\rm TIR}$-$U_h$
   Relation}\label{tab:fits}
 \footnotesize
  \begin{center}
    \begin{tabular}{ccc}
      \hline 
      & AKARI-SDSS & AKARI-HIPASS \\
      \hline
      Linear Fit$^*$ & & \\
      $a$ & $-0.097 \pm 0.006$ & $0.107 \pm 0.005$ \\
      $\sigma_{\lg U}^{\rm obs}$ & 0.354 & 0.344 \\
      $\sigma_{\lg U}^{\rm int}$ & 0.303 & 0.305 \\
      \hline
      Power-law Fit$^\dagger$ & &\\
      $a$ & $-0.085 \pm 0.006$ & $0.127 \pm 0.006$ \\
      $b$ & $0.934 \pm 0.011$  & $1.109 \pm 0.012$ \\
      $\sigma_{\lg U}^{\rm obs}$ & 0.352 & 0.350 \\
      $\sigma_{\lg U}^{\rm int}$ & 0.300 & 0.311 \\
      \hline
      3D Plane Fit$^\ddagger$ & & \\
      $a$ & $-7.566 \pm 0.108$ & $-10.341 \pm 0.120$ \\
      $b$ & $-2.179 \pm 0.034$ & $-1.850 \pm 0.035$ \\
      $c$ & $0.907 \pm 0.011$  & $1.168 \pm 0.013$ \\
      $\sigma_{\lg U}^{\rm obs}$ & 0.332 & 0.346 \\
      $\sigma_{\lg U}^{\rm int}$ & 0.277 & 0.307 \\
      \hline
      $\langle \sigma_{\lg U}^{\rm fit} \rangle$ & 0.184 & 0.160 \\
      \hline
    \end{tabular}
    \begin{minipage}{0.8\hsize}
      $^*$ $\log_{10} U_h = a + \log_{10} U_{\rm TIR}$ \\
      $^\dagger$ $\log_{10} U_h = a + b \: \log_{10} U_{\rm TIR}$ \\
      $^\ddagger$ $\log_{10} U_h = a + b \: \log_{10} 
      (R/{\rm kpc})+ c \: \log_{10} (L_{\rm TIR}/L_\odot)$ \\
    \end{minipage}
  \end{center}
\end{table}

\subsection{Dust Mass versus Metal Mass}
\label{section:dust-vs-metal}

If the interpretation of the case B2 distribution is correct for the
observed $U_{\rm TIR}$-$U_h$ correlation, this can further be tested
by checking the trend about $M_{d\rm IR}$ against $M_d$, since the
case B2 predicts the scaling of $M_{d\rm IR} / M_d \sim (\tautot)^{-1}
\propto \Sigma_d^{-1}$. For this purpose we need to measure the total
dust amount in a galaxy independently of the infrared emission.
Though it is difficult to directly measure $M_d$, the total metal mass
of a galaxy in the gas phase, $M_Z$, is an indicator of $M_d$ because
grains are made from metals.  For a part of the AKARI-SDSS sample, the
gas-phase metallicity $Z$ and $M_{\hi}$ measurements are available,
and we can estimate the \hins-gas-phase metal mass as
\begin{eqnarray}
M_{Z(\hi)} &\equiv& 1.36 \: Z \: M_{\hi} 
= 1.36 \: Z \: f_{\hi} M_{\rm H}  = f_{\hi} \: M_Z  \ ,
\end{eqnarray}
where $f_{\hi}$ is the \hi fraction of the total hydrogen gas mass,
$M_{\rm H}$.  [The numerical factor corrects the hydrogen mass
fraction of the total baryon, $X = 0.74$ (Grevesse et al. 2010).]
Some of the AKARI-SDSS galaxies and all of the AKARI-HIPASS galaxies
have $M_{\hi}$ measurements but no metallicity measurements.  To
calculate $M_{Z(\hi)}$ for these galaxies, we adopt a mean value of
$Z_{\rm av} = 0.05$ as inferred from the mass-metallicity plot of the
AKARI-SDSS galaxies (Fig. \ref{fig:M_Z}).  This metallicity is higher
than the solar abundance, but not unreasonable for dusty galaxies
selected by AKARI, taking also into account the systematic
uncertainties in the metallicity measurements (see footnote
\ref{fn:metal}).

Then the total dust mass $M_d = f_{dZ} \: M_Z$ is
related to $M_{Z(\hi)}$ as
\begin{eqnarray}
  M_d = f_{dZ} \: f_{\hi}^{-1} \: M_{Z(\hi)} \ ,
\end{eqnarray}
using the dust fraction $f_{dZ}$ of the total metal mass in
interstellar gas.  Observations of interstellar medium in our Galaxy
suggest $M_d/M_H \sim$ 0.0073, corresponding to $f_{dZ} \sim 0.3$
(Draine et al. 2007).  Although $f_{\hi}$ is unknown for our samples,
we expect $M_d/M_{Z\rm (\hi)} \sim 1$ for galaxies with $f_{\hi}$ of
order unity.

\begin{figure*}
  \begin{center}
    \includegraphics[width=14.5cm,angle=-90,scale=0.5]{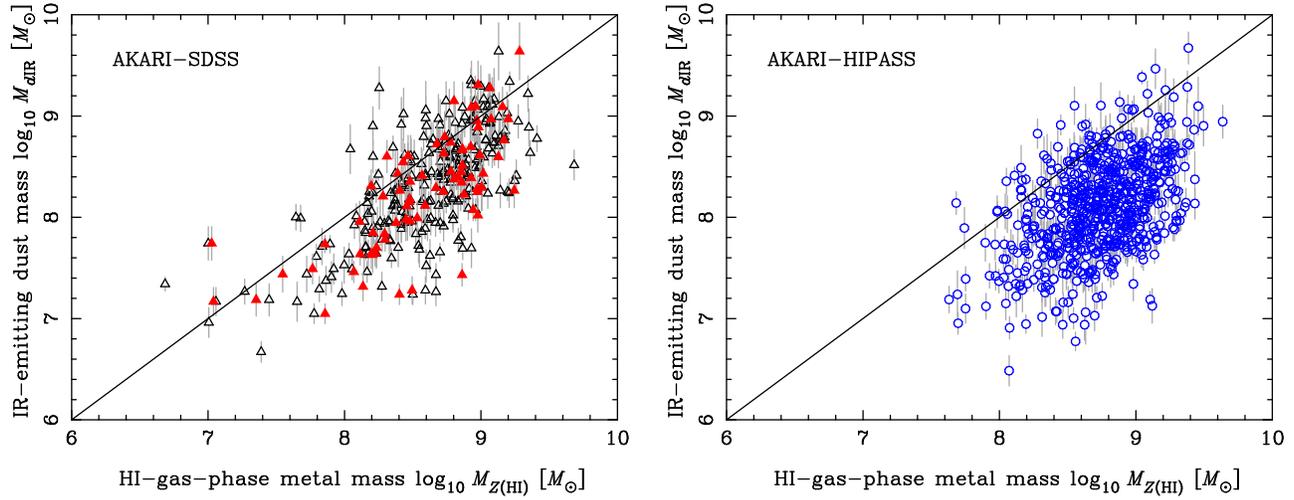}
  \end{center}
  \caption{ Heated dust mass $M_{d\rm IR}$ versus \hins-gas-phase
    metal mass $M_{Z(\hi)}$ for the AKARI-SDSS (left) and AKARI-HIPASS
    (right) samples.  For red filled triangles in the left panel,
    $M_{Z(\hi)}$ is calculated using metallicity estimates for each
    galaxy, while that of open symbols is calculated using a mean
    metallicity of $Z_{\rm av} = 0.05$ because metallicity
    measurements are not available.}
  \label{fig:M_Z_M_dIR}
\end{figure*}

Figure \ref{fig:M_Z_M_dIR} plots $M_{d\rm IR}$ against $M_{Z(\hi)}$ for
the two samples.  It can be seen that many galaxies are located at the
region of $M_{d\rm IR} < M_{Z(\hi)}$, and a clear boundary can be seen
at $M_{d\rm IR} \sim M_{Z(\hi)}$, as expected.  There are a small
number of AKARI-SDSS galaxies showing high $M_{d\rm IR}/M_{Z(\hi)}$,
but most of them are galaxies without metallicity estimates and
plotted using $Z_{\rm av}$. Most galaxies with metallicity
measurements have $M_{d\rm IR} \lesssim M_{Z(\hi)}$.

\begin{figure}
  \begin{center}
    \includegraphics[width=14cm,angle=-90,scale=0.5]{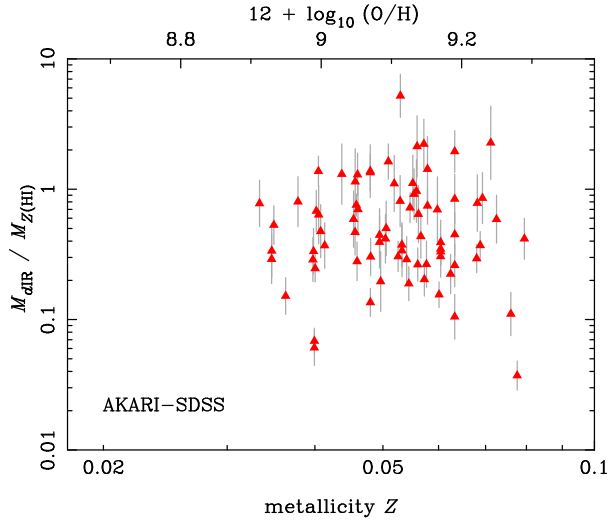}
  \end{center}
  \caption{ The ratio of $M_{d\rm IR}/M_{Z(\hi)}$ versus metallicity
    for the AKARI-SDSS galaxies with available $Z$ and $M_{\hi}$
    measurements. The errors are from those for $M_{d\rm IR}$.}
  \label{fig:Z_dZr}
\end{figure}

There are galaxies showing $M_{d\rm IR}/M_{Z(\hi)} \ll 1$, and the
origin of this small value is interesting. A possibility is that the
dust production efficiency from metals is low, i.e., $M_d \ll
M_{Z(\hi)}$.  There are observational indications that low-metallicity
dwarf galaxies have lower dust-to-metal ratios than giant galaxies
(Galliano et al. 2003, 2005; Hunt et al. 2005; but see also James et
al. 2002).  Draine et al. (2007) also found that dust-to-metal ratio
is lower for low-metallicity galaxies having $A_{\rm O} \lesssim 8.1$,
but they argued that infrared emission from such galaxies is from a
small region compared with the whole galaxy, and the dust-to-metal
ratio becomes consistent with the MW value if it is calculated using
gas mass and metallicity in the regions of infrared emission. In any
case, galaxies in our sample have $A_{\rm O} \gg 8.1$ as we showed
in Fig. \ref{fig:M_Z}, and hence the low metallicity effect is
unlikely to be the origin of the small $M_{d\rm IR}/M_{Z(\hi)}$ values
in our sample. Indeed, we plot $M_{d\rm IR}/M_{Z(\hi)}$ versus
metallicity for the AKARI-SDSS galaxies with available metallicity
measurements in Fig. \ref{fig:Z_dZr}, and there is no systematic
trend.

\begin{figure*}
  \begin{center}
    \includegraphics[width=14.5cm,angle=-90,scale=0.5]{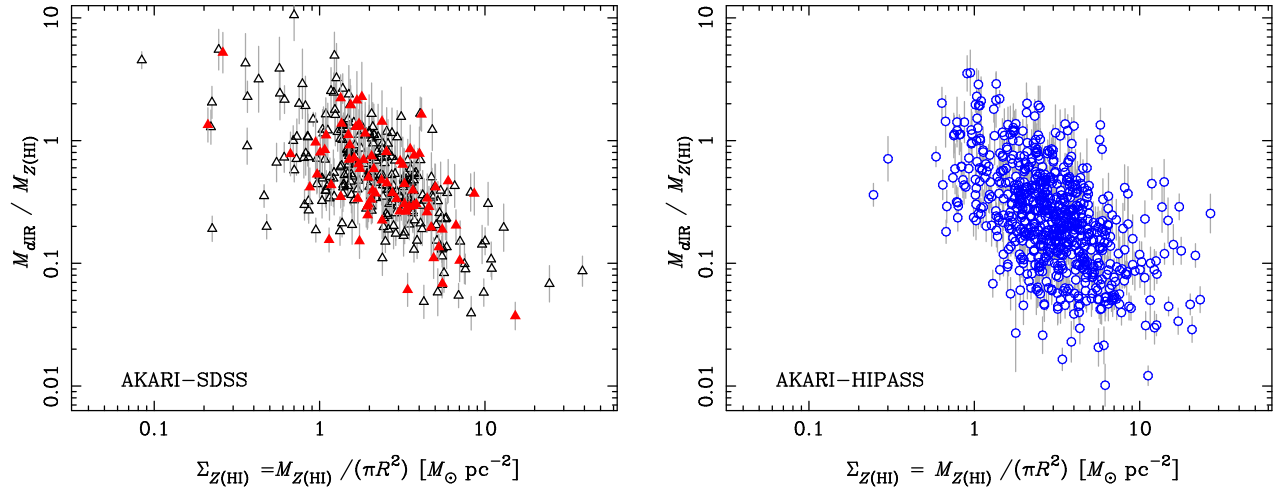}
  \end{center}
  \caption{ The ratio of $M_{d\rm IR}/M_{Z(\hi)}$ versus column
    density of \hins-gas-phase metal $\Sigma_{Z(\hi)}$ for the
    AKARI-SDSS (left) and AKARI-HIPASS (right) samples.  Different
    symbols have the same meanings as those in
    Fig. \ref{fig:M_Z_M_dIR}. The errors are from those for $M_{d\rm
      IR}$.}
  \label{fig:Sig_Z_dZr}
\end{figure*}

Instead we argue that the dispersion in $M_{d\rm IR}/M_{Z(\hi)}$
originates from the difference between $M_{d \rm IR}$ and $M_d$.  The
good correlation between $U_{\rm TIR}$ and $U_h$ indicates the case B2
distribution where only a part of dust is radiated by heating sources.
In this case we expect $M_{d\rm IR}/M_d \propto \Sigma_d^{-1}$, and
hence $M_{d\rm IR}/M_{Z(\hi)} \propto \Sigma_{Z(\hi)}^{-1}$ if $f_{dZ} \:
f_{\hi}^{-1}$ is constant for all galaxies, where $\Sigma_{Z(\hi)}
\equiv M_{Z(\hi)}/(\pi R^2)$.  Figure \ref{fig:Sig_Z_dZr} shows this
correlation, and the data of both samples clearly show the expected
correlation.  This provides a further support to the interpretation of
$U_{\rm TIR}$-$U_h$ relation by the distribution case B2.  Given the
constant nature of $\Sigma_{d\rm IR}$ found in Fig.
\ref{fig:U_TIR_Sig_Md}, the correlation $M_{d\rm IR}/M_{Z(\hi)} =
\Sigma_{d\rm IR}/\Sigma_{Z(\hi)} \propto \Sigma_{Z(\hi)}^{-1}$ may
seem rather trivial, but this figure demonstrates that $\Sigma_{d\rm
  IR}$ stays constant even if $\Sigma_{Z(\hi)}$ significantly changes.
According to this interpretation, we expect $M_{d\rm IR}/M_d \sim 1$
when $\tautot \sim 1$, i.e., $\Sigma_d \sim \Sigma_{d, \rm crit} =
0.98 \ M_\odot \: \rm pc^{-2}$.  The data is consistent with this
expectation for a reasonable value of $f_{dZ} \: f_{\hi}^{-1} \sim 1$
($\Sigma_{Z(\hi)} \sim \Sigma_d$).

Finally let us examine the assumption of $f_{dZ}f_{\hi}^{-1}$ being a
constant.  The assumption of constant $f_{dZ}$ is theoretically
reasonable if the dust production process works similarly in all
galaxies. It is also known that the dust-to-gas ratio estimated by
extinction in optical bands for nearby galaxies and quasar absorption
line systems (QALSs) is correlated well with metallicity, indicating
that $M_d/M_Z$ is roughly constant (Issa et al. 1990; Boissier et
al. 2004; Vladilo et al. 2006). It should be noted that the extinction
in QALSs is sensitive to all dust grains in a system on the line of
sight, in contrast to $M_{d\rm IR}$ that traces only dust heated by
radiation.  The variation of $f_{\hi}$ in different galaxies may also
lead to the scaling of $M_{d\rm IR}/M_{Z(\hi)} \propto
\Sigma_{Z(\hi)}^{-1}$.  However, if this effect is significant, we
expect that the sequence will appear in the region of $M_{d\rm
  IR}/M_{Z(\hi)} \gtrsim 1$ because $f_{\hi} < 1$, which is opposite
to the observed trend.

\subsection{Extension to Circumnuclear Starbursts}
\label{section:SB}

We have seen that the infrared emission from the AKARI-SDSS and
AKARI-HIPASS galaxies are described well by the case B2 of dust
distribution.  However, the dynamic range of $U_{\rm TIR}$ is limited
to $0.1 \lesssim U_{\rm TIR} \lesssim 100$, and here we examine the
AKARI-SB sample to extend this range.

\begin{figure*}
  \begin{center}
    \includegraphics[width=14.5cm,angle=-90,scale=0.9]{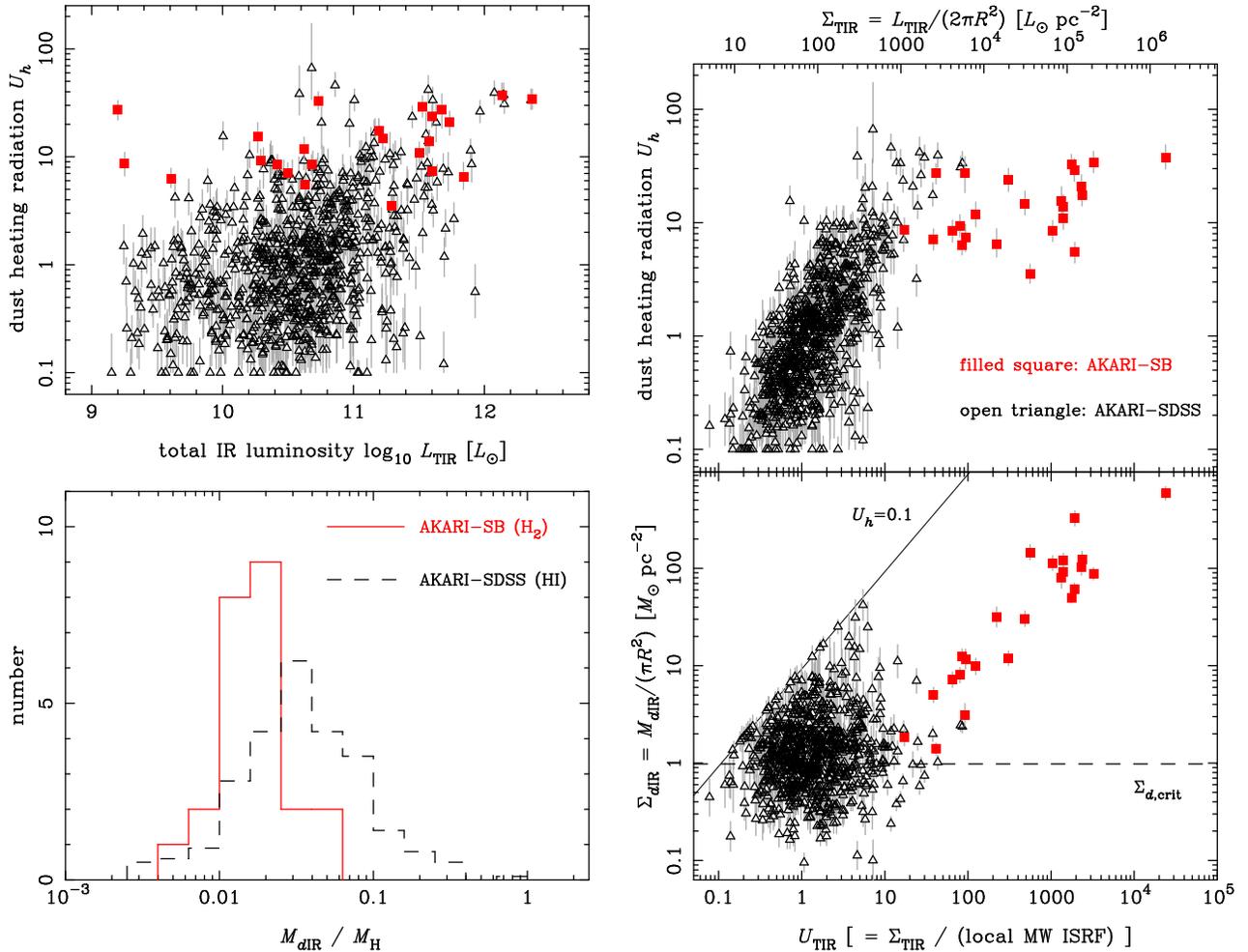}
  \end{center}
  \caption{The properties of galaxies in the AKARI-SB sample.  The
    upper-left, upper-right, and lower-right panels are the same as
    Figs.  \ref{fig:L_TIR_U_h}, \ref{fig:U_TIR_U_h}, and
    \ref{fig:U_TIR_Sig_Md}, respectively, but showing the AKARI-SB
    galaxies by the filled red squares, in comparison with the
    AKARI-SDSS sample (open triangles). The lower-left panel shows the
    histogram of the mass ratio of heated dust to hydrogen gas,
    $M_{d\rm IR}/M_{\rm H}$, where $M_{\rm H}$ is H$_2$ and \hi mass
    for the AKARI-SB and AKARI-SDSS samples, respectively.  The
    histogram of the AKARI-SDSS sample is multiplied by a factor of
    0.1 for comparison.  }
  \label{fig:sb_IR}
\end{figure*}

Figure \ref{fig:sb_IR} shows the $L_{\rm TIR}$-$U_h$, $U_{\rm
  TIR}$-$U_h$, and $U_{\rm TIR}$-$\Sigma_{d\rm IR}$ relations of the
AKARI-SB galaxies in comparison with the AKARI-SDSS sample.  The AKARI-SB
galaxies are distributed at similar locations to the AKARI-SDSS
galaxies in the $L_{\rm TIR}$-$U_h$ plane, but they have much larger
$U_{\rm TIR}$.  An interesting trend is found for $U_h$ against
$U_{\rm TIR}$; it is roughly constant and there seems an upper bound of
$U_{h, \max} \sim 50$, which is in sharp contrast to the AKARI-SDSS
galaxies showing a tight correlation of $U_h \propto U_{\rm TIR}$. But
these two data sets are not inconsistent with each other. The
dispersion of the $U_h/U_{\rm TIR}$ ratio for the AKARI-SDSS galaxies
becomes larger at $U_h \gtrsim 10$, and the distribution of the
AKARI-SDSS galaxies is smoothly connected to that of the AKARI-SB
sample. Corresponding to the constant nature of $U_h$ against $U_{\rm
  TIR}$, $\Sigma_{d\rm IR}$ is roughly proportional to $U_{\rm TIR}$,
as expected from eq. (\ref{eq:U_TIR_Sig_dIR_U_h}).

A remarkable result is obtained by comparison between $M_{d\rm IR}$
and H$_2$ mass estimated from CO observations, $M_{\rm H2}$, for the
AKARI-SB sample.  These two independently measured quantities are
correlated much better, as shown in the histogram of $M_{d\rm
  IR}/M_{\rm H2}$ in Fig. \ref{fig:sb_IR}, than the $M_{d\rm
  IR}$-$M_{\hi}$ correlation of the AKARI-SDSS galaxies. The majority
of galaxies are clustering around $M_{d\rm IR}/M_{\rm H2} \sim 0.015$
within a factor of 2, and the ratio of 0.015 is close to the
expectation of typical dust-to-hydrogen ratio $f_{d\rm H} = 1.36
f_{dZ} Z$.  It is known that molecular hydrogen is the dominant form
of hydrogen when the gas column density is very high like
circumnuclear starbursts (Kennicutt 1998; Bigiel et al. 2008).
Therefore, this result implies that the infrared-emitting dust mass is
close to the total dust mass, i.e., $M_{d\rm IR} \sim M_d$.

We consider that this result can be interpreted as follows.  As argued
by Kennicutt (1998), $L_{\rm TIR}$ can be regarded as a good SFR
indicator for the K98 SB sample. Using the relations of $M_{d\rm IR}
\sim 0.015 \: M_{\rm H2}$ and  $L_{\rm TIR} \propto M_{d\rm
  IR} \: U_h$, $U_h$ is equivalent to the star formation efficiency
(SFE), $\psi_e$, which is defined as SFR per unit hydrogen gas mass.  Using
eq. (\ref{eq:SFR_L_TIR}) and (\ref{eq:L_TIR_M_dIR_U_h}), we find
\begin{eqnarray} 
\psi_e \equiv \frac{\psi}{M_{\rm H}} = 0.24 \ U_h \ \rm [Gyr^{-1}] \ ,
\end{eqnarray} 
and the upper limit of $U_{h, \max} \sim 50$ is translated into the
upper limit of $\psi_{e, \max} \sim 10 \ \rm Gyr^{-1}$.  Then, the
systematic change of the $U_{\rm TIR}$-$U_h$ relation can be explained
if there is a fundamental upper limit on SFE by the physics of star
formation (see \S \ref{section:KS} for more discussion on the origin
of this upper limit).

If this upper limit exists, the distribution case B2 is not allowed
when $U_{\rm TIR} > U_{h, \max}$.  For the case B2, star formation
must occur within a thin layer of the disk whose dust column density
is $\Sigma_d \sim \Sigma_{d, \rm crit}$ (or $\Sigma_{\rm H} \sim
\Sigma_{\rm H, crit} \sim 10^2 \ M_\odot \rm \: pc^{-2}$, adopting a
typical dust-to-hydrogen ratio).  Then there is a maximum of the
surface SFR density [$\Sigma_{\rm SFR} \equiv \psi / (\pi R^2)$] for
the case B2 determined by $\psi_{e, \max}$ and $\Sigma_{\rm H, crit}$,
corresponding to $U_{\rm TIR} \sim U_{h, \max}$.  When $\Sigma_{\rm
  SFR}$ is higher than this, the gas column density of the star
forming layer must proportionally increase beyond $\Sigma_{\rm H,
  crit}$ because of the limit of $\psi_{e, \max}$, and hence
$\Sigma_{d\rm IR} \propto \Sigma_{\rm SFR} \propto U_{\rm TIR}$ while
$U_h$ becomes constant at $U_{h, \max}$.  This situation corresponds
to the case B1 rather than B2, and the transition from B2 occurs
around $U_{\rm TIR} \sim U_{h, \max}$.

This result should have some implications for the well-known
Kennicutt-Schmidt law on the plane of $\Sigma_{\rm H}$-$\Sigma_{\rm
  SFR}$.  We will discuss about this issue in more detail in
\S\ref{section:KS}.

\section{Discussion}
\label{section:discussion}

\subsection{Paucity of Optically Thin Galaxies and Radiative
Feedback?}
\label{section:feedback}

We have found that galaxies studied here are well described by the
distribution case B1 or B2, meaning that they are optically thick to
dust-heating radiation on a galactic scale ($\tautot \gtrsim 1$). This
is consistent with the distribution of the infrared-to-optical
luminosity ratio (Fig. \ref{fig:L_opt_L_TIR}); the distribution
extends down to $L_{\rm TIR} / L_r \sim 1$ with a sudden drop of the
galaxy number at $L_{\rm TIR} / L_r < 1$, indicating a paucity of
optically thin galaxies ($\tautot \lesssim 1$). This is also indicated
by the sequence of $M_{d\rm IR}/M_{Z(\hi)} \propto
\Sigma_{Z(\hi)}^{-1}$ in Fig. \ref{fig:Sig_Z_dZr}.  If there are a
significant number of optically thin galaxies, such galaxies should
appear in the region of $\Sigma_{Z(\hi)} \lesssim 1 \ M_\odot \rm \:
pc^{-2}$ with a constant value of $M_{d\rm IR}/M_{Z(\hi)} \sim 1$,
corresponding to $\Sigma_{d} \lesssim \Sigma_{d, \rm crit}$ and
$M_{d\rm IR}/M_d \sim 1$. However, such galaxies are not found, and
the sequence seems to stop at $\Sigma_d \sim \Sigma_{d, \rm crit}$
($\tautot \sim 1$).

\begin{figure*}
  \begin{center}
    \includegraphics[width=7cm,angle=-90,scale=0.95]{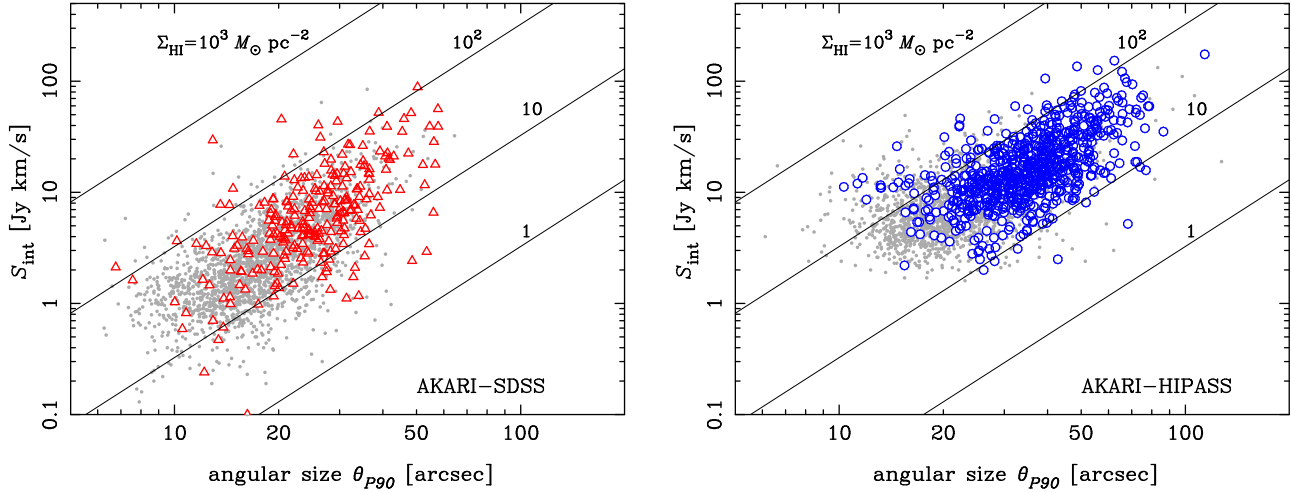}
  \end{center}
  \caption{ Velocity-integrated flux density of the 21-cm emission
    ($S_{\rm int}$) versus angular size ($\theta_{P90}$), for the
    AKARI-SDSS (left) and AKARI-HIPASS (right) samples.  The grey dots
    are for galaxies whose infrared emission was not detected by
    AKARI.  The solid lines correspond to several values of \hi column
    density $\Sigma_{\hi}$ as indicated in the panels ($S_{\rm int}
    \propto M_{\hi}/d_L^2$, $\theta_{P90} = R/d_A$, and hence
    $\Sigma_{\hi} \propto S_{\rm int}/\theta_{P90}^2$).  For these
    solid lines we assumed a redshift $z=0.02$ that is typical for the
    samples here, though the dependence on $z$ is very small only by
    the cosmological effect proportional to $(1+z)$.  }
  \label{fig:theta_S_int}
\end{figure*}

If the optical thickness to dust-heating radiation does not have any
effect on galaxy formation efficiency, a fine-tuning would be required
for the sharp drop of galaxy distribution around $\tautot \sim 1$.  A
possible explanation is a selection effect; optically thin galaxies
would have small infrared-to-optical luminosity ratio and hence could
be missed in the samples selected by infrared emission. To check this
possibility, we plot in Fig. \ref{fig:theta_S_int} velocity-integrated
21-cm flux density ($S_{\rm int}$) versus angular size
($\theta_{P90}$) for the SDSS-\hi and HIPASS samples, in comparison
with the AKARI-SDSS and AKARI-HIPASS galaxies.  The lines
corresponding to several values of \hi column density $\Sigma_{\hi}
\equiv M_{\hi}/(\pi R^2)$ are also depicted. It can be seen that there
is no significant difference between the distributions of
$\Sigma_{\hi}$ for the AKARI-detected and -nondetected galaxies, and
the paucity of galaxies in the region of $\Sigma_{\hi} \lesssim 10 \
M_\odot \: \rm pc^{-2}$ can be seen, roughly corresponding to $\tautot
\lesssim 1$ for typical values of $f_{\hi}$, $f_{dZ}$, and $Z$. Hence
the infrared selection bias cannot explain the paucity of optically
thin galaxies.  It should be noted that the beam size of HIPASS is
15.5 arcmin, which is much larger than the galaxy sizes, and hence
there should be no selection bias about surface brightness in the \hi
data.

Therefore we consider that there is a physical effect reducing the
efficiency of galaxy formation when $\tautot \lesssim 1$.  Because the
opacity is to dust-heating radiation, it is most likely the radiative
feedback by the dust photoelectric heating of ISM.  A part (typically
$\epsilon_{\rm PE} \sim$ 1--10\%) of radiation energy absorbed by dust
grains is converted to photoelectrons and subsequently heat ISM, and
this effect is large enough to be the dominant heating process in many
phases of ISM in galaxies (see, e.g., Bakes \& Tielens 1994; Wolfire
et al. 1995; Rubin et al. 2009).  Then, in $\tautot \lesssim 1$
galaxies, it is expected that the lack of self-shielding of
dust-heating radiation leads to the suppression of star formation
activity and reduction of neutral gas in ISM on a galactic scale.
Bigiel et al. (2010) found that star formation efficiency becomes
extremely low with decreasing $\Sigma_{\hi}$ in outer disk of nearby
galaxies, which is consistent with the picture discussed here.

On the other hand, in the large limit of $\Sigma_{\hi}$,
Fig. \ref{fig:theta_S_int} shows the paucity of galaxies at the region
of $\Sigma_{\hi} \gtrsim 10^2 \ M_\odot \rm \: pc^{-2}$.  This is
likely because most of gas tends to be molecular when gas column
density is high, resulting in a saturation of $\Sigma_{\hi}$, as
observed for nearby galaxies (Bigiel et al. 2008).  The efficient
self-shielding of dust-heating radiation may be responsible for
converting \hi into H$_2$.  The lack of both large and small
$\Sigma_{\hi}$ galaxies by these effects may explain the known good
correlation between H\emissiontype{I} mass and size, $M_{\hi} \propto
R^2$ (Haynes \& Giovanelli 1984; Rosenberg \& Schneider 2003;
Garcia-Appadoo et al. 2009; van der Kruit \& Freeman 2011).  The
dispersion in $M_{\hi} / R^2$ seems especially small when \hi size is
used for $R$ (Rosenberg \& Schneider 2003), further indicating that
$\tautot$ is playing an essential role.

\subsection{Implications for the Kennicutt-Schmidt Relation}
\label{section:KS}

The quantities about infrared emission discussed in this work are
mostly in surface densities of galaxies, e.g., $U_{\rm TIR}$ or
$\Sigma_{d\rm IR}$.  These quantities are closely related to the
surface density of SFR, $\Sigma_{\rm SFR}$, and surface hydrogen gas
density, $\Sigma_{\rm H}$.  Since the correlation between $\Sigma_{\rm
  SFR}$ and $\Sigma_{\rm H}$ is frequently discussed in the literature
as the Kennicutt-Schmidt relation (Schmidt 1959; Kennicutt 1998), it
is interesting to discuss about it in light of the results obtained in
this work.

\begin{figure}
  \begin{center}
    \includegraphics[width=7cm,angle=-90,scale=0.95]{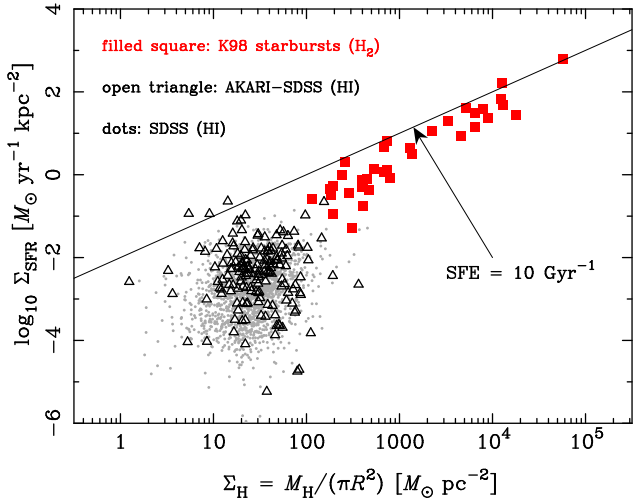}
  \end{center}
  \caption{ SFR column density versus hydrogen gas column density (the
    Kennicutt-Schmidt law). The red squares are for the circumnuclear
    starbursts of the K98 sample, estimating gas mass from H$_2$ and
    SFR from $L_{\rm TIR}$.  The AKARI-SDSS galaxies are plotted by
    triangles, and grey dots are for SDSS galaxies without AKARI
    detection, estimating gas mass from H\emissiontype{I} and SFR from
    SDSS spectra.  The relation corresponding to star formation
    efficiency (SFE) $\psi_e = 10 \ \rm Gyr^{-1}$ is shown by the
    solid line.  }
  \label{fig:sb_KS}
\end{figure}

In Figure \ref{fig:sb_KS} we show the $\Sigma_{\rm SFR}$-$\Sigma_{\rm
  H}$ plot for the AKARI-SDSS and SDSS-\hi samples, using
H\emissiontype{I} gas mass for $\Sigma_{\rm H}$.  For the K98 SB
sample, we plot the original data for the 36 galaxies in K98
($\Sigma_{\rm H}$ from H$_2$ mass by CO observations), corrected only
for SFR into the Kroupa IMF.  The SFRs of the SDSS galaxies are from
optical spectra, while those of the K98 SB sample are from $L_{\rm
  TIR}$.  There is no good correlation between $\Sigma_{\rm SFR}$ and
$\Sigma_{\hi}$ for galaxies in the SDSS-based samples at $\Sigma_{\rm
  H} \lesssim 100 \ M_\odot \: \rm pc^{-2}$, which is consistent with
the results of K98 for normal spirals and of Bigiel et al. (2008) for
sub-kpc regions of nearby galaxies. This range of $\Sigma_{\rm H}$
roughly corresponds to $\tautot \lesssim 1$, and as argued in \S
\ref{section:feedback}, the radiative feedback process seems to be
working to suppress the star formation activity.  Then the relation
between SFR and \hi gas amount would become complicated by unstable
competition between star formation and its feedback. This may give an
explanation for the large dispersion in the $\Sigma_{\rm
  SFR}$-$\Sigma_{\hi}$ relation.

On the other hand, we have found in \S\ref{section:SB} that $U_h$ is
constant at $U_h \sim 50$ for the AKARI-SB sample, corresponding to a
constant SFE of $\psi_{e, \max} \sim 10 \ \rm Gyr^{-1}$, i.e.,
$\Sigma_{\rm SFR} \propto \Sigma_{\rm H}$. This SFE is equivalent to
the star formation time scale of $t_{\rm SF, \min} \equiv \psi_{e,
  \max}^{-1} \sim 0.1$ Gyr.  It is interesting to note that this
characteristic SFE is in nice agreement with those found in the
correlation between $M_{\rm H2}$ and SFR for a wide variety of
galaxies as well as molecular clouds in the Galaxy (Gao \& Solomon
2004; Wu et al. 2005; Bigiel et al. 2008; Evans et al. 2009), when the
gas mass is measured from tracers of dense molecular gas.  These
results indicate that $\psi_{e, \max}$ is a universal characteristic
value originating from the physics of star formation in ISM, rather
than time scales on a galactic scale.  It is widely known that the
efficiency of star formation from molecular clouds is typically a few
percent in mass fraction, and typical dynamical time scales of
molecular clouds are Myr, resulting in a star formation time scale of
$\sim$ 0.1 Gyr.  This maximum SFE $\psi_{e, \max}$ seems to be achieved
on a galactic scale when $\tautot \gg 1$ ($\Sigma_{\rm H} \gg 100 \
M_\odot \: \rm pc^{-2}$) because the self-shielding effect of
dust-heating radiation is working for efficient gas cooling.
Most of interstellar gas would be molecular in such galaxies.

Now we may interpret the overall trend of the galaxy distribution on
the $\Sigma_{\rm SFR}$-$\Sigma_{\rm H}$ plane by the strong dependence
of SFE on the dust opacity, as follows.  There are very few galaxies
having $\tautot \ll 1$, because the feedback suppresses an efficient
star formation.  Galaxies appear in the region of $\tautot \gtrsim 1$
with a large dispersion in the Kennicutt-Schmidt law, because SFE
rapidly increases around $\tautot \sim 1$.  Then SFE continues to
increase and asymptotically reaches the maximum value of $\psi_{e,
  \max} \sim 10 \ \rm Gyr^{-1}$ at $\tautot \gg 1$. We expect many
galaxies around $\tautot \sim 1$ because star formation time scale is
relatively longer than that in more dusty galaxies.

\subsection{Implications for Theoretical Modelings of Galaxy Formation}
\label{section:cosmology}

There are a few implications of the results obtained by this work for
theoretical modeling of galaxy formation and evolution.  First of all,
the tight correlation between $U_h$ and $U_{\rm TIR}$ for normal
galaxies provides a convenient way to calculate infrared SEDs in
cosmological galaxy formation models, in which the total
dust-unabsorbed luminosity of stellar emission and size of a galaxy
can be calculated in a relatively straightforward manner.  When
galaxies are optically thick in terms of $\tautot$ as indicated by our
study, the total stellar luminosity gives a direct estimate of $L_{\rm
  TIR}$.  Then $U_h$ or $T_d$ can easily be calculated from the
$U_{\rm TIR}$-$U_h$ relation, which is more reliable than $L_{\rm
  TIR}$-$T_d$ relation because of the tighter correlation and more
solid physical background.  There is no particular reason to expect
that this relation does not hold for high redshift galaxies, if the
dust properties are not significantly different and the spatial
distribution of stars and dust grains is the case B2.  It must be
examined whether the existing {\it ab initio} galaxy formation models
of infrared emission satisfy this relation, especially at $z \sim 0$.

Another important implication is the strong dependence of star forming
activity on dust opacity or gas column density suggested in
\S\ref{section:feedback} and \S\ref{section:KS}.  Various feedback
processes have intensively been discussed in the context of
cosmological galaxy formation (e.g., Benson et al. 2003; Baugh 2006;
Silk 2011), but these feedback processes are not physically connected
to the galaxy-scale dust opacity, $\tautot$.  Photoionization of
hydrogens may have a similar dependence on gas column density, but it
has a much larger opacity than $\tautot$ by a factor of $\sim 3\times
10^4$ for $M_d/M_{\rm H} = 0.01$. Supernova/AGN feedbacks are unlikely
related to the galaxy-scale dust opacity.

We have argued in \S\ref{section:feedback} that the dust photoelectric
heating of ISM is the likely process responsible for the feedback
working in $\tautot \lesssim 1$ galaxies.  Using
eq. (\ref{eq:SFR_L_TIR}), 1 $M_\odot$ of star formation produces $3.2
\times 10^{49} (\epsilon_{\rm PE}/0.03)$ erg by photoelectric heating,
when most of heating radiation energy is absorbed into dust grains.
This heating energy is comparable with and may even be larger than
$1.1 \times 10^{49}$ erg produced by the supernova feedback (assuming
$10^{51}$ erg per one supernova, 8 $M_\odot$ threshold for supernova
explosions, and the Kroupa IMF).  In contrast to supernova feedback,
which is a local energy input only to dynamically connected regions,
photoelectric heating could suppress star formation on the global
scale in a galaxy when $\tautot \lesssim 1$.

Such a feedback effect as a function of dust opacity (or gas column
density) may systematically change the galaxy formation and evolution.
Cosmological objects with dark matter mass $M_{\rm DM}$ virializing at
redshift $z$ have virial density proportional to $\rho_{\rm vir}
\propto (1+z)^3$, and the available baryon mass and galaxy size
roughly scale with dark matter mass and virial radius, respectively.
Therefore we expect that gas column density scales as $\propto M_{\rm
  DM}^{1/3} (1+z)^2$, indicating that this effect will accelerate star
formation at higher redshifts, also weakly favoring star formation in
massive objects at a fixed redshift.

This may have interesting implications for some major problems in the
hierarchical galaxy formation scenario (e.g., Baugh 2006; Silk 2011).
One is the difficulty to reproduce the early appearance of massive and
quiescent galaxies at high redshifts (Cimatti et al. 2004; Glazebrook
et al. 2004), and another is that the theory predicts too large
stellar mass in massive dark halos ($\gtrsim 10^{12} M_\odot$) at $z
\sim 0$ compared with the observed galaxy luminosity function (the
so-called ``overcooling'' problem). If SFE is a strong function of
$\tautot$, it would increase the star formation activity at
high-redshift compact dark halos.  Massive objects have larger gas
column density at a fixed redshift, which may be relevant to
reproduce the well-known trend of downsizing in galaxy formation.  On
the other hand, dark halos collapsing at low redshifts are massive but
have low densities, and star formation would be suppressed to alleviate
the overcooling problem.

\subsection{Comparison with Previous Studies}

One of the most important results of this paper is the tight
correlation between the dust-heating radiation strength $U_h$ and the
galaxy-scale infrared radiation strength $U_{\rm TIR}$. The parameter
$U_{\rm TIR}$ is equivalent to the infrared surface brightness, and
the correlation of this quantity to infrared SED has been studied by
Lehnert \& Heckman (1996, hereafter LH96) and Chanial et al. (2007,
hereafter C07).

LH96 estimated the size of 32 nearby starburst galaxies by H$\alpha$
images, and found a correlation between $L_{\rm TIR}/(\pi R_{\rm
  H\alpha})^2$ and infrared color temperature estimated by 60/100
$\mu$m flux ratios. They argued that a surrounding dust screen
radiated by a radiation field strength of $L_{\rm TIR}/(\pi R_{\rm
  H\alpha})^2$ is an adequate zeroth-order description of their data,
which is equivalent to our interpretation of the case B2 for normal
star-forming galaxies.  However, we also found that $U_h$ becomes
roughly constant for circumnuclear starbursts in the high SFR density
region. The LH96 galaxies are in the range of $10^3 \lesssim L_{\rm
  TIR}/(\pi R_{\rm H\alpha}^2) \lesssim 10^5 \ L_\odot\: \rm pc^{-2}$,
roughly corresponding to $10 \lesssim U_{\rm TIR} \lesssim 10^3$
assuming that the H$\alpha$ radius is not different from the $r$-band
radius used here.  This is the region of transition from $U_h \propto
U_{\rm TIR}$ to a constant $U_h$, and indeed, the distribution of the
LH96 galaxies is flatter than expected from the simple surrounding
screen in the $S_{\rm 60\mu m}/S_{\rm 100\mu m}$ versus $L_{\rm
  TIR}/(\pi R_{\rm H\alpha}^2)$ plane (see their Fig. 5).  Our result
is thus not inconsistent with that of LH96, but revealed the
correlation more quantitatively by larger statistics, using not only
starbursts but also normal galaxies.

C07 studied on samples ($\sim$400 galaxies in total) including
non-starburst galaxies.  They estimated galaxy sizes by radio
continuum (RC) images, and found a correlation of $\Sigma_{\rm TIR} =
L_{\rm TIR} / (2 \pi R_{\rm RC}^2) \propto T_d^\delta$ with $\delta
\sim 18.5$, where $T_d$ is the dust temperature estimated by assuming
the modified blackbody spectrum. The range of $\Sigma_{\rm TIR}$ is
$\sim 10^1$--$10^5 L_\odot \: \rm pc^{-2}$ ($U_{\rm TIR} \sim
0.1$--$10^3$).  In contrast to LH96, they concluded that the
surrounding dust screen is ruled out because the observed slope of the
correlation is significantly different from the expectation ($\delta =
4 + \beta$).  Instead, they proposed another interpretation based on
the Kennicutt-Schmidt law.  Assuming a constant dust-to-gas ratio
($\Sigma_{d\rm IR} \propto \Sigma_{\rm gas}$), $L_{\rm TIR} \propto
\psi$, and the original form of the Kennicutt-Schmidt law
($\Sigma_{\rm SFR} \propto \Sigma_{\rm gas}^{1.4}$, Kennicutt 1998),
the basic equation of $L_{\rm TIR} \propto M_{d\rm IR} \:
T_d^{4+\beta}$ can be reduced to $\Sigma_{\rm TIR} \propto
T_d^{3.5(4+\beta)}$, which is close to the observed $\delta$ value
with the emissivity index of $\beta \sim 1$--2.

Our result of $U_{\rm TIR} \propto U_h \propto T_d^{4+\beta}$ for
normal galaxies ($U_{\rm TIR} \lesssim 30$) is equivalent to $\delta =
4 + \beta$, while the constant $U_h$ for the circumnuclear starbursts
($U_{\rm TIR} \gtrsim 30$) corresponds to $\delta = \infty$.  The
$\delta$ value obtained by C07 may be a result of the fit including
the transition region of $U_{\rm TIR}$ between the two regimes.  Their
interpretation based on the Kennicutt-Schmidt law is not supported
from our study, because we have seen that SFR is not proportional to
$L_{\rm TIR}$, and $\Sigma_{d\rm IR}$ is roughly constant against
change of $\Sigma_{\rm gas}$, for normal star-forming galaxies.  There
is a large scatter in the Kennicutt-Schmidt law, and it is unlikely
that the tight correlation between $U_h$ and $U_{\rm TIR}$ emerges
from this law.

Besides the larger statistics, the strength of our study is that we
estimated the dust-heating radiation field strength by fitting
realistic physical dust models to the four AKARI/FIS photometric bands
at 65--160 $\mu$m covering the broad thermal SED peak.  It is good to
use $U_h$ instead of $T_d$ to examine the correlation against $U_{\rm
  TIR}$, because they have the same physical dimensions and hence not
only the scaling between the two but also the ratio $U_h/U_{\rm TIR}$
is physically meaningful.  It should also be noted that $T_d$ is
meaningful only for large grains in thermal equilibrium. Another
important source of difference is that, as mentioned in
\S\ref{section:L_TIR_U_h}, the AKARI-based samples include galaxies
with smaller $U_h$ than the IRAS-based samples used in the previous
studies, which enabled us to clearly find the linear correlation between
$U_h$ and $U_{\rm TIR}$ in the normal galaxy regime.

\section{Summary and Conclusions}
\label{section:summary}

We have constructed two new low-redshift galaxy samples based on the
AKARI/FIS All Sky Bright Source Catalog: the AKARI-SDSS sample (878
galaxies) cross-correlated with the SDSS galaxies and available \hi
data, and the AKARI-HIPASS sample (711 galaxies) cross-correlated with
the HIPASS \hi galaxy catalog. To complement these for wider dynamic
range of surface SFR density, the AKARI-SB sample (24 galaxies) has
also been constructed from nearby circumnuclear starbursts.  These
samples include only galaxies having high quality AKARI/FIS fluxes at
least in two bands to ensure the reliability of SED fittings.  

We then studied these samples to understand the physics determining
galactic-scale infrared SEDs, especially the broad thermal peak by
large grains. The strength of dust-heating radiation field ($U_h$,
roughly corresponding to the dust temperature as $U_h \propto
T_d^{5.75}$), heated dust mass ($M_{d\rm IR}$), and total infrared
luminosity ($L_{\rm TIR}$) were derived from the fittings of physical
dust models to the observed infrared SEDs, and their correlations with
various physical quantities [optical size and luminosity, SFR
($\psi$), stellar mass ($M_*$), metallicity, and \hi or H$_2$ gas
masses] were investigated.

We found a tight power-law correlation between the $L_{\rm TIR}/\psi$
ratio and the specific SFR (SSFR, $\psi_s \equiv \psi/M_*$), as
$L_{\rm TIR}/\psi \propto \psi_s^{-0.54}$, which is valid except for
very dusty galaxies having a large infrared-to-optical luminosity
ratio (\S \ref{section:L_TIR_L_opt_SFR}).  Therefore SSFR is a good
measure of the contribution from young stars to $L_{\rm TIR}$.  This
relation may be useful to estimate the third parameter when two of
SFR, $M_*$, and $L_{\rm TIR}$ are known.  The widely used value of
$L_{\rm TIR}/\psi$ as a SFR indicator is valid only for galaxies
having largest SSFR of $\psi_s \: \gtrsim \: 0.1 \: \rm Gyr^{-1}$, and
relatively aged stellar population must be taken into account in the
dust-heating energy budget for most of the AKARI-SDSS galaxies.

The most important result of this work is the discovery of the tight
linear correlation between $U_h$ and galactic-scale mean infrared
radiation field $U_{\rm TIR} \propto L_{\rm TIR}/(\pi R^2)$ for the
AKARI-SDSS and AKARI-HIPASS samples, taking the $r$-band optical size
as $R$ (\S \ref{section:U_TIR_U_h}).  The ratio of $U_h/U_{\rm TIR}$
is close to unity in the same physical units, and the physical
dispersion along this relation in $U_h$ is about 0.3 dex ($\sim 13$\%
in the dust temperature).  This $U_{\rm TIR}$-$U_h$ relation is nicely
explained by a thin disk of dust-heating sources embedded in a
thicker, optically-thick ($\tautot \gtrsim 1$) dust disk (the case B2
in Fig. \ref{fig:dust_dist_schem}), where $\tautot$ is the effective
dust opacity averaged over wavelength with the weight of the SED of
the interstellar radiation field (ISRF).  This also means that
infrared SEDs of galaxies are determined mainly by ISRF on a galactic
scale, rather than those on smaller scales like individual star
forming regions or molecular clouds.

The $U_{\rm TIR}$-$U_h$ relation is particularly useful in theoretical
modeling of infrared emission, giving a simple method to determine
dust SED from $L_{\rm TIR}$ and $R$ that can be predicted relatively
easily in cosmological galaxy formation models.  It may also be useful
in observational studies; e.g., when flux of dust emission is
available only in one band for a galaxy but its size is known, one can
make a guess about the SED shape and $L_{\rm TIR}$ by searching for $U_h$
that satisfies the relation of $U_{\rm TIR} \sim U_h$.

Since only dust in the layer of one optical depth is heated, the dust
column density estimated by infrared emission should be constant at
$\Sigma_{d\rm IR} \sim \Sigma_{d, \rm crit}$ corresponding to
$\tau_{d, \rm eff} = 1$, which was also confirmed by our data.  This
interpretation predicts that the heated fraction of the total dust
mass ($M_{d\rm IR}/M_d$) scales with the total dust column density as
$\propto \Sigma_d^{-1}$, and the observed trends of $M_{d\rm
  IR}/M_{Z(\hi)} \propto \Sigma_{Z(\hi)}^{-1}$ and $M_{d\rm
  IR}/M_{\hi} \propto \Sigma_{\hi}^{-1}$ support this expectation,
where $Z(\hi)$ denotes the metal in \hi gas phase (\S
\ref{section:dust-vs-metal}).  An important implication is that, for
many normal galaxies, the dust mass derived from infrared emission
gives only a lower bound, with a significant amount of cold dust not
contributing to infrared emission.

On the other hand, the data of the AKARI-SB sample indicate that there
is an upper limit of $U_h \lesssim 50$ (in units of the local ISRF
around the solar neighbourhood), and $U_h$ becomes constant at this
value for intensive starbursts having $U_{\rm TIR} \gtrsim 50$ (\S
\ref{section:SB}).  The H$_2$ mass is tightly correlated with the
heated dust mass $M_{d\rm IR}$, and the upper limit of $U_h$
corresponds to the maximum star formation efficiency (SFE, defined as
SFR per hydrogen gas mass) of $\psi_{e, \max} \sim 10 \ \rm
Gyr^{-1}$. This maximum SFE likely originates from the physics of star
formation in ISM, and the spatial distribution of heating sources and
dust changes from the case B2 to B1 (the same mixed distribution for
stars and dust) in Fig. \ref{fig:dust_dist_schem} around $U_{\rm TIR}
\sim 50$.

All the galaxies studied here seem optically thick to dust heating
radiation. The distribution of $\tautot$ extends down to $\tautot \sim
1$, but there is a sharp drop of the number of optically thin galaxies
at $\tautot \lesssim 1$. We argued in \S \ref{section:feedback} that
this is unlikely to be a result of selection effects, but there seems
a feedback effect to suppress the formation of optically thin
galaxies.  Because it is related to dust opacity, the feedback is
likely by the photoelectric heating of ISM by dust grains. This
process is known to be the major heating in many phases of ISM, and
the self-shielding of dust-heating radiation on a galactic scale may
be a necessary condition to avoid a global suppression of star
formation. It is often stated that starburst galaxies are dusty, but
if this hypothesis is true, one may rather state that they are
starbursts {\it because} they are dusty.

We discussed implications of our results for galaxy formation in
general. The strong dependence of galactic-scale gas cooling on
$\tautot$ may be responsible for the known correlation between \hi
mass and size ($M_{\hi} \propto R^2$), which is on the line
corresponding to $\tautot \sim 1$ (\S \ref{section:feedback}). The
Kennicutt-Schmidt law (SFR surface density $\Sigma_{\rm SFR}$ versus
hydrogen gas surface density $\Sigma_{\rm H}$) may be understood by
the maximum SFE $\psi_{e, \max}$ as the baseline with the strong
dependence of SFE on $\tautot$ by the feedback (\S \ref{section:KS}).
The feedback by dust opacity may also have interesting
implications for galaxy formation in the cosmological context, because
it would accelerate star formation in massive objects at
high redshifts. Such an effect may be helpful to solve some of the
problems currently discussed in the field of cosmological galaxy
formation (\S \ref{section:cosmology}).

It is quite intriguing to examine whether the scaling relations found
in this work, especially $U_{\rm TIR}$-$U_h$, hold also for high
redshift galaxies.  Planned infrared/submillimeter facilities will
give such opportunities in the near future. Especially, ALMA will be a
quite powerful facility for this purpose by the unprecedented flux
sensitivity and angular resolution at wavelengths covering the
thermal SED peak at high redshifts.

The electronic catalog data of the samples constructed in this
work are available on request to the authors. 


\bigskip

We would like to thank Y. Inoue for useful discussions.  This work is
based on observations with AKARI, a JAXA project with the
participation of ESA.  This work was supported in part by the
Grant-in-Aid for Scientific Research (19740099 for TT, 20740105 and
23340046 for TTT, 22740123 for MN, and 22111506 for all) from the
Ministry of Education, Culture, Sports, Science and Technology (MEXT).
TT and RM were supported by the Global COE Program ``The Next
Generation of Physics, Spun from Universality and Emergence'' from
MEXT.  TTT was supported by the Global COE Program ``Quest for
Fundamental Principles in the Universe: from Particles to the Solar
System and the Cosmos'' from MEXT.  TTT was also supported by the
Program for Improvement of Research Environment for Young Researchers
from the Special Coordination Funds for Promoting Science and
Technology.  MARK was supported by the Research Fellowship for Young
Scientists from the Japan Society for the Promotion of Science (JSPS).


\end{document}